\documentclass[12pt,  draftclsnofoot,onecolumn,article,romanappendices]{IEEEtran}
\usepackage{setspace}

\usepackage{cite}
\ifCLASSINFOpdf
   \usepackage[pdftex]{graphicx}
\else

  \usepackage[dvips]{graphicx}
\fi
\usepackage[cmex10]{amsmath}
\usepackage{amssymb}
\usepackage{amsfonts}
\usepackage{amsthm}
\usepackage{array}
\usepackage{dsfont}
\usepackage{amsbsy}
\usepackage{epstopdf}
\usepackage{graphicx,color}
\usepackage{hyperref}

\newtheorem{proposition}{Proposition}
\newtheorem{remark}{Remark}
\newtheorem{corollary}{Corollary}
\long\def\symbolfootnote[#1]#2{\begingroup%
\def\thefootnote{\fnsymbol{footnote}}\footnote[#1]{#2}\endgroup}
\newtheorem{theorem}{Theorem}
\newtheorem{definition}{Definition}

\usepackage{amsmath,amssymb,amsthm,mathrsfs,amsfonts,dsfont,stmaryrd}
\usepackage{mathtools}
\usepackage{color}
\usepackage{graphicx} 
\usepackage{subfig} 
\allowdisplaybreaks[4]

\newcommand{\dv}{\mathbf} 
\newcommand{\mc}{\mathcal} 
\newcommand{\mkv}{-\!\!\!\!\minuso\!\!\!\!-}

\usepackage{algorithm} 
\usepackage{algpseudocode}

\algnewcommand{\Inputs}[1]{%
  \State \textbf{Inputs:}
  \Statex \hspace*{\algorithmicindent}\parbox[t]{.8\linewidth}{\raggedright #1}
}
\algnewcommand{\Initialize}[1]{%
  \State \textbf{initialization}
  \Statex \hspace*{\algorithmicindent}\parbox[t]{.95\linewidth}{\raggedright #1}
}

\begin{document}

\title{In-network Compression for Multiterminal Cascade MIMO Systems}

\author{I\~naki Estella Aguerri \qquad \qquad Abdellatif Zaidi\\

\thanks{
This work will be presented in part at the IEEE Int'l Conference on Communications 2017 \cite{Estella:ICC:17:Chained}.
I. Estella Aguerri is with the Mathematical and Algorithmic Sciences Lab, 
France Research Center, 92100 Boulogne-Billancourt, France. A. Zaidi was with Universit\'e Paris-Est, France, and is currently on leave at the Mathematical and Algorithmic Sciences Laboratory, Huawei France Research Center, 92100 Boulogne-Billancourt, France.
  Email: \{\tt inaki.estella@huawei.com, abdellatif.zaidi@u-pem.fr\}.}
}

\maketitle
\begin{abstract} 
We study the problem of receive beamforming in uplink cascade multiple-input multiple-output (MIMO) systems as an instance of that of cascade multiterminal source coding for lossy function computation. Using this connection, we develop two coding schemes for the second and show that their application leads to beamforming schemes for the first. In the first coding scheme, each terminal in the cascade sends a description of the source that it observes; the decoder reconstructs all sources, lossily, and then computes an estimate of the desired function. This scheme improves upon standard routing in that every terminal only compresses the innovation of its source w.r.t. the descriptions that are sent by the previous terminals in the cascade. In the second scheme, the desired function is computed gradually in the cascade network, and each terminal sends a finer description of it. In the context of uplink cascade MIMO systems, the application of these two schemes leads to \textit{centralized} receive-beamforming and \textit{distributed} receive-beamforming, respectively. Numerical results illustrate the performance of the proposed methods and show that they outperform standard routing.
\end{abstract}

\IEEEpeerreviewmaketitle

\vspace{-4mm}
\section{Introduction}

\begin{figure}
\centering 
\includegraphics[width=0.60\textwidth]{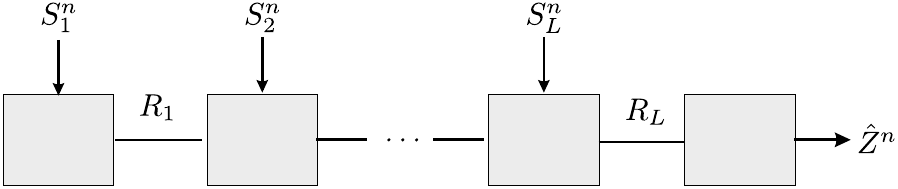}
\caption{ Multi-terminal cascade source  coding for lossy function computation.}
\label{SystemModel}
\vspace{-7mm}
\end{figure}

Consider the cascade communication system for function computation shown in Figure~\ref{SystemModel}. Terminal $l$, $l=1,\hdots,L$, observes, or measures, a discrete memoryless source $S^n_l$ and communicates with Terminal $(l+1)$ over an error-free finite-capacity link of rate $R_l$. Terminal $(L+1)$ does not observe any source, and plays the role of a decoder which wishes to reconstruct a function $Z^n$ lossily, to within some average fidelity level $D$, where $Z_i=\varphi(S_{1,i},\hdots,S_{L,i})$ for some function $\varphi(\cdot)$.  The memoryless sources $(S^n_1, \hdots, S^n_L)$ are arbitrary correlated among them, with joint measure $p_{S_1,\hdots,S_L}(s_1,\hdots,s_L)$. For this communication system, optimal tradeoffs among compression rate tuples $(R_1,\hdots,R_L)$ and allowed distortion level $D$, captured by the rate-distortion region of the model, are not known in general, even if the sources are independent. For some special cases, inner and outer bounds on the rate-distortion region, that do not agree in general, are known, e.g., in \cite{Cuff2009_Cascade} for the case $L=2$. A related work for the case $L=2$ has also appeared in~\cite{Permu2012_CascTriang}. For the general case with $L \geq 2$, although a single-letter characterization of the rate-distortion region seems to be out of reach, one can distinguish essentially two different transmission approaches or modes. In the first mode, each terminal operates essentially as a \textit{routing} node. That is,  each terminal in the cascade sends an appropriate compressed version, or description, of the source that it observes; the decoder reconstructs all sources, lossily, and then computes an estimate of the desired function. In this approach, the computation is performed \textit{centrally}, at only the decoder, i.e., Terminal $(L+1)$. In the second mode, Terminal $l$, $l=1,\hdots,L$, \textit{processes} the information that it gets from the previous terminal, and then describes it, jointly with its own observation or source, to the next terminal. That is, in a sense, the computation is performed \textit{distributively} in the network. (See, e.g., \cite{Sefidgaran2016DistFunComp,  Park2015Multihop, Gover2016ISIT}, where variants of this approach are sometimes referred to as \textit{in-network processing}).
 
\begin{figure}
\centering 
\includegraphics[width=0.65\textwidth]{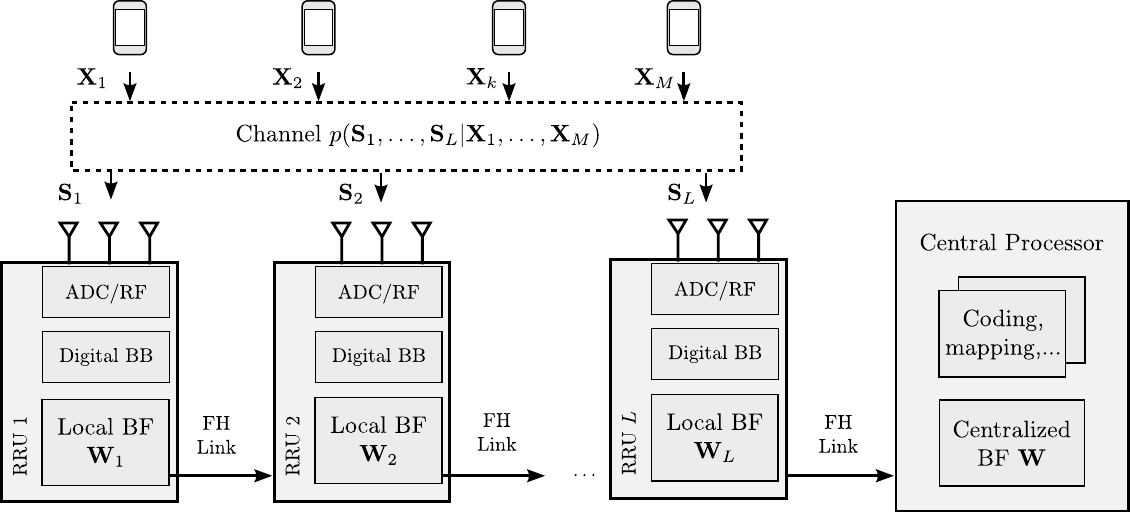}
\caption{Chained MIMO architecture for uplink Massive MIMO systems.}
\label{fig:ChainedMIMO}
\vspace{-7mm}
\end{figure}

Consider now the seemingly unrelated uplink multiple-input multiple-output (MIMO) system model shown in Figure~\ref{fig:ChainedMIMO}. In this model, $M$ users communicate concurrently with a common base station (BS), as in standard uplink wireless systems. The base station is equipped with a large number of  antennas, e.g., a Massive MIMO BS; and the baseband processing is distributed across a number, say $L$, of modules or radio remote units (RRUs). The modules are connected each to a small number of antennas; and are concatenated in a line network, through a common fronthaul link that connects them to a central processor (CP) unit. This architecture, sometimes referred to as ``chained MIMO"~\cite{Puglielli:2015ICCW:ScalableMassive} and proposed as an alternative to the standard one in which each RRU has its dedicated fronthaul link to the CP \cite{Somekh:2007:IT, DelCoso:2009:TWir, Sanderovich:2009:IT, Hwan:2013:VT,  Park:2013:SPLett, Zhou:2014:JSAC, Nazer:ISIT:2009,HongCaire:IT:2013, Estella2016CQCF2, Estella:Allerton20015}, offers a number of advantages and an additional degree of flexibility if more antennas/modules are to be added to the system. The reader may refer to \cite{Shepard:2012:APM:2348543.2348553, Shepard:2013:AFM:2500423.2505302, Viera2014FlexibleTest, Balan:2013:USE:2491246.2491254} where examples of testbed implementations of this novel architecture can be found. For this architecture, depending on the amount of available channel state information (CSI), receive-beamforming operations may be better performed centrally at the CP or distributively across RRUs. Roughly, if CSI is available only at the CP, not at the RRUs, it seems reasonable that beamforming operations be performed only centrally, at the CP. In this case, RRU $l$, $l=1,\hdots,L$, sends a compressed version $\hat{\dv S}_l$ of its received signal $\dv S_l$ to the CP which first collects the vector $(\hat{\dv S}_l,\hdots,\hat{\dv S}_L)$ and then performs receive-beamforming on it. In contrast, if local CSI is available or can be acquired at the RRUs, due to the linearity of the receive beamforming (which is a simple matrix multiplication) parts of the receive beamforming operations can be performed distributively at the RRUs (see Section III).   

The above shows some connections among the model of Figure~\ref{fig:ChainedMIMO} and that, more general, of Figure~\ref{SystemModel}. In this paper, we study them using a common framework. Specifically, we develop two coding schemes for the multiterminal cascade source coding problem of Figure~\ref{SystemModel}; and then show that their application to the uplink cascade MIMO system of Figure~\ref{SystemModel} leads to schemes for receive-beamforming which, depending on the amount of available CSI at the RRUs, are better performed centrally at the CP or distributively across RRUs. In the first coding scheme, each terminal in the cascade sends a description of the source that it observes; the decoder reconstructs all sources lossily and then computes an estimate of the desired function. This scheme improves upon standard routing in that every terminal only compresses the innovation of its source w.r.t. the descriptions that are sent by the previous terminals in the cascade. In the second scheme, the desired function is computed gradually in the cascade network; and each terminal sends a finer description of it. Furthermore, we also derive a lower bound on the minimum distortion at which the desired function can be reconstructed at the decoder by relating the problem to the Wyner-Ziv type system studied in \cite{Yamamoto:IT:1982}. Numerical results show that the proposed methods outperform standard compression strategies and perform close to the lower bound in some regimes.  


 \vspace{-3mm}
\subsection{Notation}

Throughout, we use the following notation. Upper case letters are used to denote random variables, e.g., $X$; lower case letters used to denote realizations of random variables $x$; and calligraphic letters denote sets, e.g., $\mathcal{X}$.  The cardinality of a set $\mc X$ is denoted by $|\mc X|$. The length-$n$ sequence $(X_1,\ldots,X_n)$ is denoted as  $X^n$; and, for $(1\leq k \leq j \leq n)$ the sub-sequence $(X_k,X_{k+1},\ldots, X_j)$ is denoted as  $X_{k}^j$. Boldface upper case letters denote vectors or matrices, e.g., $\dv X$, where context should make the distinction clear. For an integer $L \geq 1$, we denote the set of integers smaller or equal $L$ as $\mathcal{L}\triangleq \{1,\ldots, L\}$; and, for $1 \leq l \leq L$, we use the shorthand notations $\mathcal{L}_l\triangleq\{ 1,2,\ldots l\}$, $\mathcal{L}_l^c\triangleq\{ l+1\ldots L\} $ and $\mathcal{L}/l\triangleq \{1,\ldots, l-1,l+1,\ldots,L\}$. We denote the covariance of a vector $\mathbf{X}$ by $\mathbf{\Sigma}_{\mathbf{x}}\triangleq\mathrm{E}[\mathbf{XX}^H]$;  $\mathbf{\Sigma}_{\mathbf{x},\mathbf{y}}$ is the cross-correlation  $\mathbf{\Sigma}_{\mathbf{x},\mathbf{y}}\triangleq \mathrm{E}[\mathbf{XY}^H]$, and the conditional correlation matrix of $\mathbf{X}$ given $\mathbf{Y}$ as $\mathbf{\Sigma}_{\mathbf{x}|\mathbf{y}}\triangleq \mathbf{\Sigma}_{\mathbf{x}}-\mathbf{\Sigma}_{\mathbf{x},\mathbf{y}}\mathbf{\Sigma}_{\mathbf{y}}^{-1}\mathbf{\Sigma}_{\mathbf{y},\mathbf{x}}$. The length-$N$ vector with all entries equal zero but the $l$-th element which is equal unity is denoted as $\boldsymbol\delta_{l}$, i.e., $\boldsymbol\delta_{l}\triangleq[\mathbf{0}_{1\times l-1},1,\mathbf{0}_{1\times N-l}]$; and the $l{\times}N$ matrix whose entries are all zeros, but the first $l$ diagonal elements which are equal unity, is denoted by $\mathbf{\bar{I}}_l$, $[\mathbf{\bar{I}}_l]_{i,i}  = 1$  for $i\leq l$ and $0$ otherwise. We also, define $\log^+(\cdot)\triangleq\max\{0,\log(\cdot)\}$.

 \vspace{-2mm}

\section{Cascade Source Coding System Model}

Let $\{S_{1,i},S_{2,i},\ldots,S_{L,i}\}_{i=1}^{n} = (S_1^n,\ldots,S_{L}^n)$ be a sequence of $n$ independent and identically distributed (i.i.d.) samples of the $L$-dimensional source $(S_{1},S_{2},\ldots,S_{L})$  jointly distributed as $p(s_1,\ldots,s_L)$ over $\mathcal{S}_1\times \ldots \times\mathcal{S}_L$. For convenience, we denote $S_{L+1}^n\triangleq \emptyset$.

A cascade of $(L+1)$ terminals are concatenated as shown in Figure \ref{SystemModel}, such that Terminal $l$, $l=1,\ldots,L$,  is connected to Terminal $(l+1)$ over an error-free link of capacity $R_l$ bits per channel use. Terminal $(L+1)$ is interested in reconstructing a sequence $Z^n$ lossily, to within some fidelity level, where $Z_i= \varphi(S_{1,i},\ldots,S_{L,i})$, $i=1,\ldots, n$, for some function  $\varphi:\mathcal{S}_1\times\ldots \times \mathcal{S}_L\rightarrow \mathcal{Z}$. To this end, Terminal $l$, $l=1,\ldots,L$, which observes the sequence $S_l^n$ and receives message $m_{l-1}\in\mathcal{M}_{l-1}\triangleq \{1,\ldots, M_{l-1}^{(n)}\}$ from Terminal $(l-1)$, generates a message $m_{l}\in\mathcal{M}_{l}$ as $m_l=f_l^{(n)}(s^n_l, m_{l-1})$ for some encoding function $f_l^{(n)}:\mathcal{S}_l^n\times \mathcal{M}_{l-1}\rightarrow \mathcal{M}_l$,  and forwards it over the error-free link of capacity $R_l$ to Terminal $(l+1)$. 
At Terminal $(L+1)$, the message $m_L$ is mapped to an estimate $\hat{Z}^n=g^{(n)}(m_L)$ of $Z^n$, using some mapping $g^{(n)}:\mathcal{M}_{L}\rightarrow\mathcal{\hat{Z}}^n$.
 Let $\mathcal{\hat{Z}}$ be the reconstruction alphabet and $d:\mathcal{Z}\times \mathcal{\hat{Z}}\rightarrow [0,\infty)$ be a single letter distortion. The distortion between $Z^n$ and the reconstruction  $\hat{Z}^n$ is defined as 
$d(z^n;\hat{z}^n)=\frac{1}{n}\sum_{i=1}^nd(z_i;\hat{z}_i)$.

\begin{definition}
A tuple $(R_1,\ldots, R_{L})$ is said to achieve distortion $D$ for the cascade multi-terminal source coding problem if there exist $L$ encoding functions  $f_l^{(n)}:\mathcal{S}_l\times \mathcal{M}_{l-1}\rightarrow \mathcal{M}_l$, $l=1,\ldots, L$, and a  function $g^{(n)}:\mathcal{M}_{L}\rightarrow\mathcal{\hat{Z}}$ such that 
\begin{align}
R_{l}&\geq \frac{1}{n}\log M_l^{(n)}, \; l=1,\ldots, L\; \text{ and }\;
D\geq \frac{1}{n}\mathrm{E}[d(Z^n;\hat{Z}^n)].\nonumber
\end{align} 

\end{definition}
The rate-distortion (RD) region $\mathcal{R}(D)$ of the cascade multi-terminal source coding problem is defined as the closure of all rate tuples $(R_1,\ldots, R_{L})$  that achieve distortion $D$.

\section{Schemes for Cascade Source Coding} \label{sec:achievability}

In this section, we develop two coding schemes for the cascade source coding model of Figure~\ref{SystemModel} and analyze the RD regions that they achieve.

\subsection{Improved Routing (IR)}\label{ssec:InnoRouting}

A simple strategy which is inspired by standard routing (SR) in graphical networks and referred to as multiplex-and-forward in \cite{Park2015Multihop} has Terminal $l$, $l=1,\hdots,L$, forward a compressed version of its source to the next terminal, in addition to the bit stream received from the previous terminal in the cascade (without processing). The decoder decompresses all sources and then outputs an estimate of the desired function. In SR, observations are compressed independently and correlation with the observation of the next terminal in the cascade is not exploited. 

In this section, we propose a scheme, to which we refer to as ``Improved Routing'' (IR), which improves upon SR by compressing at each terminal its observed signal $S_l^n$ into a description $U_l^n$ considering the compressed observations from the previous terminals, i.e., $(U_1^n,\ldots, U_{l-1}^n)$ as side information available both at the encoder and the decoder \cite{elGamal:book}. Thus, each terminal only compresses the innovative part of the observation with respect to the compressed signals from previous terminals (see Section \ref{ssec:CentBF_ImpRout}). In doing so, it uses $B_l$ bits per source sample. Along with the produced compression index of rate $B_l$, each terminal also forwards the bit stream received from the previous terminal to the next one without processing. The decoder successively decompresses all sources and outputs an estimate of the function of interest. 

\vspace{-2mm}
\begin{theorem}\label{th:InnoRout}
The RD region  $\mathcal{R}_{\mathrm{IR}}(D)$ that is achievable with the IR scheme is given by the union of rate tuples $(R_1,\ldots, R_{L})$ satisfying 
\begin{align}
R_l\geq \sum_{i=1}^{l} I(S_{i};U_{i}|U_{1},\dots,U_{i-1}), \label{eq:RoutConst} \quad \text{for} \quad l=1,\ldots, L,
\end{align}
for some joint pmf  $p(s_1,\ldots, s_L)\prod_{l=1}^{L} p(u_l|s_l,u_1,\ldots,u_{l-1})$ and function $g$, s.t.  $\hat{Z} = g(U_1,\dots, U_L)$ and
$\mathrm{E}[d(Z,\hat{Z})]\leq D$.
\end{theorem}
\vspace{-6mm}
\begin{remark}
The auxiliary random variables $(U_1,\hdots,U_L)$ that are involved in \eqref{eq:RoutConst} satisfy the following Markov Chains
\vspace{-2mm}
\begin{equation}
U_l \mkv (S_l,U_{\mc L_{l-1}}) \mkv (S_{\mc L/l},U_{\mc L_l^c}) \quad \text{for} \quad l=1,\hdots,L,
\end{equation}
where 
$U_{\mc L_{l-1}} = (U_1,\ldots,U_{l-1})$ and $U_{\mc L_l^c} = (U_{l+1},\ldots,U_{L})$.\qed
\end{remark}

\noindent\textbf{Outline Proof:}
Fix $\epsilon >0$, and a joint pmf $p(s_1,\ldots,s_L,u_1,\ldots, u_L)$ that factorizes as 
\vspace{-2mm}
\begin{align}
p(s_1,\ldots,s_L,u_1,\ldots, u_L) = p(s_1,\ldots, s_L)\prod_{l=1}^{L} p(u_l|s_l,u_1,\ldots,u_{l-1}),
\end{align}
 and a reconstruction function  $g(\cdot)$ such that $\mathrm{E}[d(Z;g(U_1,\ldots, U_L))]\leq D/(1+\epsilon)$. Also, fix non-negative $R_1,\ldots, R_L$ such that for $R_l\geq R_1+\cdots + R_{l-1}$, for $l=1,\ldots, L$.

\noindent\textit{Codebook generation:} 
Let $B_l = R_l - (R_1 + \cdots + R_{l-1})$ for $l=1,\ldots, L$. Generate a codebook $\mathcal{C}_{1}$ consisting of a collection of $2^{nB_1}$ codewords  $\{u_1^n(i_1)\}$, indexed with $i_1 =1,\ldots, 2^{nB_1}$, where codeword $u_1^n(i_1)$ has its elements generated  i.i.d. according to $p(u_1)$. 
For each index $i_1$, generate a codebook $\mathcal{C}_{2}(i_1)$ consisting of a collection of $2^{n B_2}$ codewords  $\{u_2^n(i_1,i_2)\}$ indexed with $i_2 =1,\ldots, 2^{nB_2}$, where codeword  $u_2^n(i_1,i_2)$ is generated independently and i.i.d. according to $p(u_2|u_1)$. Similarly, for each index tuple $(i_1,\ldots, i_{l-1})$ generate a codebook $\mathcal{C}_l(i_1,\ldots, i_{l-1})$ of $2^{nB_{l}}$ codewords $\{u_l^n(i_1,\ldots,i_l)\}$ indexed with $i_l =1,\ldots, 2^{nB_l}$, and where codeword $u_l^n(i_1,\ldots,i_l)$ has its elements generated i.i.d. according to $p(u_l|u_1,\ldots,u_{l-1})$.

\noindent\textit{Encoding at Terminal 1:} Terminal $1$ finds an index $i_1$ such that $u^n_{1}(i_1)\in \mathcal{C}_{1}$ is strongly $\epsilon$-jointly typical with $s_{1}^n$, i.e., $(u_1^n(i_1),s^n_1)\in\mathcal{T}^{(n)}_{[U_1S_1]}$.
Using standard arguments, this step can be seen to have vanishing probability of error as long as $n$ is large enough and 
\begin{align}\label{eq:IR_cond_00}
B_1\geq I(S_1;U_1).
\end{align}
Then, it forwards $m_1 = i_1$ to Terminal 2.

\noindent \textit{Encoding at Terminal $l\geq 2$: } Upon reception of $m_{l-1}$ with the indices $m_{l-1}=(i_1,\ldots, i_{l-1})$, Terminal $l$ finds an index $i_l$  such that $(u_1^n(i_1),\ldots,u_{l}^n(i_1,\ldots,i_l))$  are strongly $\epsilon$-jointly typical with $s_{l}^n$, i.e., $(u_1^n(i_1),\ldots,u_{l}^n(i_1,\ldots,i_l),s_l^n)\in \mathcal{T}^{(n)}_{[U_1,\ldots,U_l,S_l]}$. 
Using standard arguments, this step can be seen to have vanishing probability of error as long as $n$ is large enough and 
\begin{align}\label{eq:IR_cond_0}
B_l\geq I(U_l;S_l|U_1,\ldots,U_{l-1}).
\end{align}

Then, it forwards $i_l$ and $m_{l-1}$ to Terminal $(l+1)$ as $m_l=(i_l,m_{l-1})$.

\noindent\textit{Reconstruction at end Terminal $(L+1)$:} Terminal $(L+1)$ collects all received indices as $m_{L} = (i_1,\ldots, i_L)$, and reconstructs the codewords $(u_1^n(i_1),\ldots,u_{L}^n(i_1,\ldots,i_L))$. Then, it reconstructs an estimate of $Z^n$ sample-wise as $\hat{Z}_i = g(u_{1,i}(i_1),u_{2,i}(i_1,i_2),\ldots,u_{L,i}(i_1,\ldots,i_L))$, $i=1,\ldots, n$. Note that in doing so, the average distortion constraint is satisfied.

\noindent Finally, substituting $B_l = R_l - (R_1 + \cdots + R_{l-1})$ in \eqref{eq:IR_cond_00} and \eqref{eq:IR_cond_0}, we get  \eqref{eq:RoutConst}. This completes the proof of Theorem \ref{th:InnoRout}.
\qed

\begin{remark}\label{rm:RemarkWZ}
In the coding scheme of Theorem~\ref{th:InnoRout},  the compression rate on the communication hop between Terminal $l$ and Terminal $(l+1)$, $l=1,\ldots, L$, can be improved further (i.e., reduced)  by taking into account sequence $S^n_{l+1}$ as decoder side information, through Wyner-Ziv binning. The resulting strategy, however, is not of ``routing type'', unless  every Wyner-Ziv code is restricted to account for the worst side information ahead in the cascade, i.e., binning at Terminal $l$ accounts for the worst quality side information among the sequences $\{S^n_j, j=l+1,..,L\}$. Also, in this case, since the end Terminal $(L+1)$, or CP,  does not observe any side information, i.e., $ S^n_{L+1}=\emptyset$, this strategy makes most sense if the Wyner-Ziv codes are chosen such that the last relay terminal in the cascade, i.e., Terminal $L$, recovers an estimate of the desired function and then sends it using a standard rate-distortion code to the CP in a manner that allows the latter to reconstruct the desired function to within the desired fidelity level.
\end{remark}

The above routing scheme necessitates that every terminal $l$, $l=1,\hdots,L$, reads the compressed bit streams from previous terminals in the cascade prior to the compression of its own source. This is reflected through $(U_1,\hdots,U_{l-1})$ treated not only as decoder side information but also as encoder side information. From a practical viewpoint, treating previous terminals streams as encoder side information improves rates but generally entails additional delays. The following corollary specializes the result of Theorem~\ref{th:InnoRout} to the case in which $(U_1,\hdots,U_{l-1})$ is treated only as decoder side information, i.e., the auxiliary random variables are restricted to satisfy that $U_l \mkv S_l \mkv (U_1,\hdots,U_{l-1})$ forms a Markov chain. We also present an alternate coding scheme that is based on successive Wyner-Ziv coding~\cite{WZ76}. 

\begin{corollary}\label{th:WZR}
The RD region $\mathcal{R}_{\mathrm{WZR}}(D)$ that is achievable with the WZR scheme is given by the set of rate tuples $(R_1,\ldots, R_{L})$ satisfying 
\begin{equation}
R_l\geq \sum_{i=1}^{l} I(S_{i};U_{i}|U_{1},\dots,U_{i-1}),\label{eq:rateConst_WZ} \quad \text{for} \quad l=1,\ldots, L,
\end{equation}
for some joint pmf $p(s_{1},\ldots,s_L)\prod_{l=1}^{L} p(u_l|s_l)$ and function $g$ s.t.  $\mathrm{E}[d(Z,g(U_1,\dots, U_L))] \leq D$.
\end{corollary}

\begin{remark}
The auxiliary random variables $(U_1,\hdots,U_L)$ that are involved in \eqref{eq:rateConst_WZ} satisfy the following Markov Chains
\vspace{-3mm}
\begin{equation}
U_l \mkv S_l \mkv (S_{\mc L/l},U_{\mc L/l}), \quad \text{for} \quad l=1,\hdots,L,
\end{equation}
\vspace{-3mm}
where $S_{\mc L/l} = (S_1,\ldots,S_{l-1},S_{l+1},\hdots,S_L)$ and $U_{\mc L/l} = (U_1,\ldots,U_{l-1},U_{l+1},\hdots,U_L)$.
\qed
\end{remark}
\noindent\textbf{Outline Proof:}
The proof of Corollary~\ref{th:WZR} follows by applying successively standard Wyner-Ziv source coding~\cite{WZ76}. Hereafter, we only outline the main steps, for the sake of brevity. Fix $\epsilon >0$ and a joint pmf $p(s_1,\hdots,s_L,u_1,\hdots,u_L)$ that factorizes as 
\vspace{-3mm}
\begin{equation}
p(s_1,\hdots,s_L,u_1,\hdots,u_L) = p(s_{1},\ldots,s_L)\prod_{l=1}^{L} p(u_l|s_l)
\end{equation}
and a function $g(\cdot)$ such that $\mathrm{E}[d(Z,g(U_1,\dots, U_L))] \leq D/(1+\epsilon)$. 
Also, fix non-negative $R_1, \hdots, R_L$, such that for $R_l \geq R_1+\cdots +R_{l-1}$ for $l=1,\hdots, L$.
 
\noindent\textit{Codebook generation:} Let non-negative $\hat{R}_1, \hdots, \hat{R}_L$, and set $B_l=R_l-(R_1+\cdots +R_{l-1})$ for $l=1,\hdots, L$. Generate $L$ codebooks $\{\mathcal{C}_{l}\}$, $l=1,\ldots, L$, with codebook $\mathcal{C}_{l}$ consisting of a collection of $2^{n(B_l+\hat{R}_l)}$ independent codewords $\{u_l^n(i_l)\}$ indexed with $i_l=1,\hdots,2^{n(B_l+\hat{R}_l)}$, where codeword $u_l^n(i_l)$ has its elements generated i.i.d. according to $p(u_l)$. Randomly and independently assign these codewords into $2^{n B_l}$ bins $\{\mathcal{B}_{j_l}\}$ indexed with $j_l = 1,\hdots, 2^{nB_l}$, each containing $2^{n\hat{R}_l}$ codewords. 

\noindent\textit{Encoding at Terminal $l$:} Terminal $l$ finds an index $i_l$ such that $u^n_{l}(i_l)\in \mathcal{C}_{l}$ is strongly $\epsilon$-jointly typical\footnote{For formal definitions of strongly $\epsilon$-joint typicality, the reader may refer to \cite{elGamal:book}.} with $s_{l}^n$, i.e., $(u_l^n(i_l),s^n_l)\in\mathcal{T}^{(n)}_{[U_lS_l]}$.  Using standard arguments, it is easy to see that this can be accomplished with vanishing probability of error as long as $n$
is large and 
\begin{align}\label{eq:CovConditions_WZR}
B_l+\hat{R}_l \geq I(S_{l};U_l).
\end{align}
Let $j_l$ such that  $\mathcal{B}_{j_l}\ni u^n_l(i_l)$. Terminal $l$ then forwards the bin index $j_l$ and the received message $m_{l-1} = (j_1,\ldots, j_{l-1})$ to Terminal $l$ as $m_l = (m_{l-1}, j_l)$.

\noindent\textit{Reconstruction at the end Terminal $(L+1)$:} Terminal $(L+1)$ collects all received bin indices as $m_L = (j_1,\ldots, j_L)$, and reconstructs the codewords $u_1^n(i_1),\ldots, u_L^n(i_L)$ successively in this order, as follows. Assuming that codewords $(u_{1}^n(i_1),\ldots,u_{l-1}^n(i_{l-1}))$ have been reconstructed correctly, it finds the appropriate codeword $u_l^n(i_l)$ by looking in the bin $\mathcal{B}_{j_l}$ for the unique $u_l^n(i_l)$ that is $\epsilon$-jointly typical with $(u_{1}^n(i_1),\ldots,u_{l-1}^n(i_{l-1}))$. Using standard arguments, it is easy to see that the error in this step has vanishing probability  as long as $n$ is large and 
\begin{align}\label{eq:MKConditions_proc_WZR}
\hat{R}_l< I(U_{l}; U_{1},\ldots,U_{l-1}).
\end{align}
\noindent Terminal $(L+1)$ reconstructs an estimate of $Z^n$ sample-wise as $\hat{Z}_i = g(u_{1,i}(i_1),u_{2,i}(i_2),\ldots,u_{L,i}(i_L))$, $i=1,\ldots, n$. Note that in doing so the average distortion constraint is satisfied.

\noindent Finally, substituting $B_l=R_l-(R_1+\cdots +R_{l-1})$ and combining \eqref{eq:CovConditions_WZR} and \eqref{eq:MKConditions_proc_WZR}, we get \eqref{eq:rateConst_WZ}; and this completes the proof of Corollary~\ref{th:WZR}.
\qed

\begin{remark}
Note that the rate constraints in Theorem~\ref{th:InnoRout}  and  Corollary~\ref{th:WZR} are identical. However, $\mathcal{R}_{\mathrm{WZR}}(D)\subseteq \mathcal{R}_{\mathrm{IR}}(D)$ since the set of feasible pmfs in IR is larger than that in WZR.
\end{remark}
\vspace{-5mm}
\begin{figure}
\centering 
\includegraphics[width=0.45\textwidth]{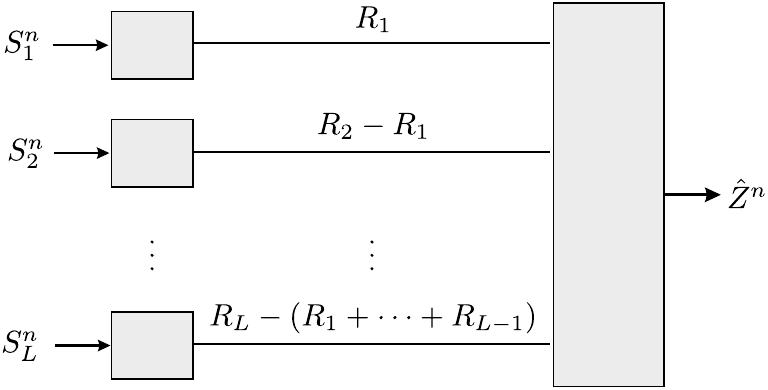}
\caption{Distributed source coding model for the routing scheme of Corollary~\ref{th:WZR}.}
\label{fig:BergerTungModel}
\vspace{-5mm}
\end{figure}
\begin{remark}
As it can be conveyed from the proof of Corollary~\ref{th:WZR}, since every Terminal $l$, $l=1,\hdots,L$, uses a part $(R_1+\hdots+R_{l-1})$ of its per-sample rate $R_l$ to simply route the bit streams received from the previous terminals in the cascade and the remaining per-sample $B_l=R_l-(R_1+\hdots+R_{l-1})$ bits to convey a description of its observed source $S^n_l$, the resulting scheme can be seen as one for the model shown in Figure~\ref{fig:BergerTungModel} in which the terminals are connected through parallel links to the CP.
Using this connection, the performance of the above WZR scheme can be further improved by compressing the observations \`a-la Berger-Tung~\cite{Berger1989MultiterminalSourceEncoding}. 
\end{remark}
\begin{remark}
In accordance with Remark \ref{rm:RemarkWZ}, for the model of Figure \ref{SystemModel} yet another natural coding strategy is one in which one decomposes the problem into $L$ successive Wyner-Ziv type problems for function computation, one for each hop.  Specifically, in this strategy one sees the communication between Terminal $l$ and Terminal $(l+1)$, $l=1,\ldots, L$, as a Wyner-Ziv source coding problem with two-sided state information, state information $S^n_l$ at the encoder and state information $S^n_{l+1}$ at the decoder. This strategy, which is not of ``routing type'', is developed in the next section.
\end{remark}

\subsection{In-Network Processing (IP)}\label{ssec:InlineProcessing}

In the routing schemes in Section  \ref{ssec:InnoRouting}, the function of interest is computed at the destination from the compressed observations, i.e.,  the terminals have to share the fronthaul to send a compressed version of their observations to Terminal $(L+1)$. We present a scheme to which we refer to as ``In-Network Processing'' (IP), in which instead, each terminal computes a part of the function to reconstruct at the decoder so that the function of interest is computed along the cascade. To that end, each terminal decompresses the signal received from the previous terminal and jointly compresses it with its observation to generate an estimate of the part of the function of interest, which is forwarded to the next terminal (see Section \ref{ssec:InlineBeam}). Correlation between the computed part of the function and the source at the next terminal $S^n_{l+1}$ is exploited through Wyner-Ziv coding.  Note that by decompressing and recompressing the observations at each terminal, additional distortion is introduced \cite{Cuff2009_Cascade}. 

\begin{theorem}\label{th:inline}
The RD region $\mathcal{R}_{\mathrm{IP}}(D)$ that is achievable with IP is given by the union of rate tuples $(R_1,\ldots, R_{L})$ satisfying  
\vspace{-2mm}
\begin{align}\label{eq:cond_inline}
R_l\geq I(S_{l},U_{l-1};U_{l}|S_{l+1}),\quad l=1,\ldots, L,
\end{align}
for some joint pmf $p(s_{1},\ldots,s_L)\prod_{l=1}^L p(u_l|s_l,u_{l-1})$ and a function $g$, such that $\hat{Z} = g(U_L)$ and
$\mathrm{E}[d(Z,\hat{Z})]\leq D$.
\end{theorem}

\begin{remark}
The auxiliary random variables $(U_1,\hdots,U_L)$ that are involved in \eqref{eq:cond_inline} satisfy the following Markov Chains
\vspace{-2mm}
\begin{equation}
U_l \mkv (S_l,U_{l-1}) \mkv (S_{\mc L/l},U_{\mc L/\{l-1,l\}}) \quad \text{for} \quad l=1,\hdots,L,
\end{equation}
where $S_{\mc L/l} = (S_1,\ldots,S_{l-1},S_{l+1},\hdots,S_L)$, $U_{\mc L/{\{l-1,l\}}} = (U_1,\ldots,U_{l-2},U_{l+1},\hdots,U_L)$.
\qed
\end{remark}

\noindent \textbf{Outline Proof:}
Fix $\epsilon >0$ and a joint pmf $p(s_1,\ldots, s_L,u_1,\ldots, u_L)$ that factorizes as
\begin{align}
p(s_1,\ldots, s_L,u_1,\ldots, u_L) = p(s_{1},\ldots,s_L)\prod_{l=1}^L p(u_l|s_l,u_{l-1}),
\end{align}
and a function $g(\cdot)$ such that $\mathrm{E}[d(Z;g(U_L))]\leq D/(1+\epsilon) $. Also fix non-negative $R_1,\ldots, R_L$.

\noindent\textit{Codebook generation:} Let non-negative $\hat{R}_1,\ldots, \hat{R}_L$. Generate $L$ codebooks $\{\mathcal{C}_{l}\}$, $l=1,\ldots, L$, with codebook $\mathcal{C}_{l}$ consisting of a collection of $2^{n(R_l+\hat{R}_l)}$ independent codewords $\{u_l^n(i_l)\}$, indexed with $i_l=1,\ldots,2^{n(R_l+\hat{R}_l)}$, where codeword $u_l^n(i_l)$ has its elements generated randomly and independently i.i.d. according to $p(u_l)$. Randomly and independently assign these codewords into $2^{nR_l}$ bins $\{\mathcal{B}_{j_l}\}$ indexed with $j_l = 1,\ldots, 2^{nR_l}$, each containing $2^{n\hat{R}_l}$ codewords.

\noindent\textit{Encoding at Terminal $1$}: Terminal $1$ finds an index $i_1$ such that $u_1^n\in \mathcal{C}_1$ is strongly $\epsilon$-jointly typical with $s_1^n$, i.e., $(u_1^n(i_1),s_1^n)\in\mathcal{T}^{(n)}_{[U_1S_1]}$. Using standard arguments, it is easy to see that this can be accomplished with vanishing probability of error as long as $n$ is large and 
\vspace{-1mm}
\begin{align}\label{eq:MKConditions_proc_0}
R_1+\hat{R}_1\geq I(S_1;U_1).
\end{align}

Let $j_1$ such that $\mathcal{B}_{j_1}\ni u_1^n(i_1)$. Terminal $1$ then forwards the index $j_1$ to Terminal $2$.

\noindent\textit{Decompression and encoding at Terminal $l\geq 2$:}
 Upon reception of the bin index $m_{l-1}=j_{l-1}$ from Terminal $(l-1)$, Terminal $l$ finds $u_{l-1}^n(i_{l-1})$ by looking in the bin $\mathcal{B}_{j_l}$ for the the unique $u_{l-1}^n(i_l)$ that is $\epsilon$-jointly typical with $s_{l}^n$.
Using standard arguments, it can be seen that this can be accomplished with vanishing probability of error as long as $n$ is large enough and
\begin{align}\label{eq:MKConditions_proc_1}
\hat{R}_{l-1}< I(U_{l-1}; S_{l}).
\end{align}
\noindent Then, Terminal $l$ finds an index $i_l$ such that $u_l^n(i_l)\in \mathcal{C}_{l}$ is strongly $\epsilon$-jointly typical with $(s_l^n,u_{l-1}^n(i_{l-1}))$, i.e., $(s_l^n,u^n_{l-1}(i_{l-1}),u^n_{l}(i_l))\in\mathcal{T}^{(n)}_{[SU_{l-1}U_l]}$.  Using standard arguments, it can be seen that this can be accomplished with vanishing probability of error as long as $n$ is large and 
\vspace{-1mm}
\begin{align}\label{eq:MKConditions_proc_2}
R_l+\hat{R}_l\geq I(S_l,U_{l-1};U_l).
\end{align}
\vspace{-1.5mm}
\noindent Let $j_l$ such that $\mathcal{B}_{j_l}\ni u_l^n(i_l)$. Terminal $l$ forwards the bin index to Terminal $(l+1)$ as $m_{l} = j_l$.
 \noindent\textit{Reconstruction at end Terminal $(L+1)$:} Terminal $(L+1)$ collects the bin index $m_{L} = j_L$ and reconstructs the codeword $u_{L}^n(i_L)$ by looking in the bin $\mathcal{B}_{j_L}$. Since Terminal $(L+1)$ does not have available any side information sequence, from \eqref{eq:MKConditions_proc_1}, successful recovery of the unique $u_{L}^n(i_L)$ in the bin $\mathcal{B}_{j_L}$  requires $\hat{R}_L = 0$. That is, each bin contains a single codeword and $j_L=i_L$. Then, Terminal $(L+1)$ reconstructs an estimate of $Z^n$ sample-wise as $\hat{Z}_i = g(u_{L,i}(i_L))$, $i=1,\ldots, n$. In doing so, the average distortion constraint is satisfied.

\noindent Finally, combining \eqref{eq:MKConditions_proc_0}, \eqref{eq:MKConditions_proc_1} and \eqref{eq:MKConditions_proc_2}, we get \eqref{eq:cond_inline}. This completes the proof of Theorem \ref{th:inline}.
\qed
\begin{remark}
It is shown in \cite{Cuff2009_Cascade} that for $L=2$, in general none of the IR and IP schemes outperform the other; and a scheme combining the two strategies is proposed.
\end{remark}

\section{Centralized and Distributed Beamforming in Chained MIMO Systems}\label{sec:Beamforming}
In this section, we apply the cascade source coding model to study the achievable  distortion in a Gaussian uplink MIMO system with a chained MIMO architecture (C-MIMO) in which $M$ single antenna users transmit over a Gaussian channel to $L$ RRUs as shown in Figure \ref{fig:ChainedMIMO}.
The signal received at RRU $l$, $l= 1,\ldots, L$, equipped with $K$ antennas, $\mathbf{S}_l\in \mathds{C}^{K\times 1}$, is given by
\begin{align}
\mathbf{S}_l = \mathbf{H}_l\mathbf{X}+\mathbf{N}_l,
\end{align}
 where $\mathbf{X} = [X_1,\ldots, X_M]^T$  is the signal transmitted by the $M$ users and $X_m\in\mathds{C}$, $m=1,\ldots, M$ is the signal transmitted by user $m$. We assume that each user satisfies an average power constraint $\mathrm{E}[|X_m|^2]\leq \mathrm{snr}$, $m=1,\ldots, M$, where $\mathrm{snr}>0$; $\mathbf{H}_l\in\mathds{C}^{K\times M}$ is the channel between the $M$ users and RRU $l$ and $\mathbf{N}_l\!\in\!\mathds{C}^{K\times 1}\!\sim\!\mathcal{CN}(\mathbf{0},\mathbf{I})$ is the additive ambient noise.

The transmitted signal by the $M$ users is assumed to be distributed as $\mathbf{X} \sim\mathcal{CN}(\mathbf{0},\mathrm{snr}\mathbf{I})$ and we denote the observations at the $L$ RRUs as $\mathbf{S}=[\mathbf{S}_1^T,\ldots, \mathbf{S}_L^T]^T$. Thus, we have $\mathbf{S}\sim\mathcal{CN}(\mathbf{0},\mathbf{\Sigma}_{\mathbf{s}})$, where $\mathbf{\Sigma}_{\mathbf{s}}=\mathrm{snr}\mathbf{H}_{\mathcal{L}}\mathbf{H}_{\mathcal{L}}^H+\mathbf{I}$ and $\mathbf{H}_{\mathcal{L}}\triangleq [\mathbf{H}_1^T,\ldots,\mathbf{H}_L^T]^T$. 

In traditional receive-beamforming, a beamforming filter $\mathbf{W}\in\mathds{C}^{M\times L\cdot K}$ is applied  at the decoder on the received signal $\mathbf{S}$ to estimate the channel input $\mathbf{X}$ with the linear function
\begin{align}\label{eq:Beamforming}
\mathbf{Z}\triangleq \mathbf{W}\mathbf{S}. 
\end{align} 

In C-MIMO, the decoder (the CP) is interested in computing the receive beamforming signal $\eqref{eq:Beamforming}$ with minimum distortion, although $\mathbf{S}$ is not directly available at the CP but remotely observed at the terminals. Depending on the available CSI, receive-beamforming computation may be better performed centrally at the CP or distributively across the RRUs: 

\textit{Centralized Beamforming: }If CSI is available only at the CP, not at the RRUs, it seems reasonable that beamforming operations are performed only centrally at the CP. In this case, RRU $l$, $l=1,\hdots,L$, sends a compressed version $\hat{\dv S}_l$ of its output signal $\dv S_l$ to the CP, which first collects the vector $(\hat{\dv S}_l,\hdots,\hat{\dv S}_L)$, and then performs receive-beamforming on it. 

\textit{Distributed Beamforming: } If local CSI is available at the RRUs, or can be acquired, receive beamforming operations can be performed distributively along the cascade. Due to linearity the joint beamforming operation \eqref{eq:Beamforming} can be expressed as a function of the received source as 
\begin{align}\label{eq:function}
\mathbf{Z}= \mathbf{W}\mathbf{S} =  \mathbf{W}_1 \mathbf{S}_1+\cdots+\mathbf{W}_L\mathbf{S}_{L},
\end{align}
where $\mathbf{W}_l\in \mathds{C}^{M \times K}$ corresponds to blocks of $K$ columns of $\mathbf{W}$ such that $[\mathbf{W}_1,\ldots,\mathbf{W}_L ] =\mathbf{W}$.  In this case, the receive beamforming signal can be computed gradually in the cascade network, by letting the RRUs compute a part of the desired function, e.g., as proposed in Section \ref{ssec:InlineBeam}, RRU $l$, $l=1,\hdots,L$ computes an estimate of $\mathbf{W}_1 \mathbf{S}_1+\cdots+\mathbf{W}_l\mathbf{S}_{l}$.

The distortion between $\mathbf{Z}$ and the reconstruction of the beamforming signal $\mathbf{\hat{Z}}$ at the CP is measured with the sum-distortion 
\begin{align}
d(\mathbf{Z},\mathbf{\hat{Z}})\triangleq \mathrm{Tr}\{(\mathbf{Z}-\mathbf{\hat{Z}})(\mathbf{Z}-\mathbf{\hat{Z}})^H\}. \label{eq:sumdist}
\end{align}

For a given fronthaul tuple $(R_1,\ldots, R_{L})$ in the RD region $\mathcal{R}(D)$, the minimum achievable average distortion $D$  is characterized by the distortion-rate function\footnote{This formulation is equivalent to the rate-distortion framework considered in Section \ref{sec:achievability}; here we consider the distortion-rate formulation for convenience.} given by
\begin{align}
D(R_1,\ldots,R_L) \triangleq \min\{D\geq 0: (R_1,\ldots, R_L)\in \mathcal{R}(D)\}.
\end{align} 

Next, we study the distortion-rate function in a Gaussian C-MIMO model under centralized and distributed beamforming with the schemes proposed for the cascade source coding problem.

\subsection{Centralized Beamforming with Improved Routing} \label{ssec:CentBF_ImpRout}

In this section, we consider distortion-rate function of the IR scheme in Section \ref{ssec:InnoRouting} applied for centralized beamforming. Each RRU forwards a compressed version of the observation to the CP, which estimates the receive-beamforming signal $\mathbf{Z}$ from the decompressed observations. While the optimal test channels are in general unknown, next theorem gives the distortion-rate function of IR for centralized beamforming for the C-MIMO setup under jointly distributed Gaussian test channels.

\begin{theorem}\label{lem:InnoGaus}
The distortion-rate function for the IR scheme under jointly Gaussian test channels is given by 
\begin{align}\label{eq:InnoGaus}
D_{\mathrm{IR}}(R_1,\ldots,R_L)=&\min_{\mathbf{K}_1,\ldots,\mathbf{K}_L}
\mathrm{Tr}\{\mathbf{\Sigma}_{\mathbf{z}}-\mathbf{\Sigma}_{\mathbf{z},\mathbf{u}_{\mathcal{L}}}\mathbf{\Sigma}_{\mathbf{u}_{\mathcal{L} }}^{-1}\mathbf{\Sigma}_{\mathbf{z},\mathbf{u}_{\mathcal{L}}}^H \}\\
&\text{s.t. }R_l\geq B_1+\ldots+B_l,\quad  l=1,\ldots, L, \label{eq:FHConstraints}\\
&\quad \; \; B_l\triangleq \log|\mathbf{\Sigma}_{\mathbf{s}_l|\mathbf{u}_{\mathcal{L}_{l-1} }}+\mathbf{K}_l|/|\mathbf{K}_l|,
\end{align}
where $\mathbf{\Sigma}_{\mathbf{s}_l|\mathbf{u}_{\mathcal{L}_{l-1} }} = \mathbf{\Sigma}_{\mathbf{s}_l}- \mathbf{\Sigma}_{\mathbf{s}_l,\mathbf{u}_{\mathcal{L}_{l-1} }}\mathbf{\Sigma}_{\mathbf{u}_{\mathcal{L}_{l-1} }}^{-1} \mathbf{\Sigma}_{\mathbf{s}_l,\mathbf{u}_{\mathcal{L}_{l-1} }}^H
$ and
%
$\mathbf{\Sigma}_{\mathbf{s}_l} = \boldsymbol{\delta}_l\mathbf{\Sigma}_{\mathbf{s}}\boldsymbol{\delta}_l^T$, 
$\mathbf{\Sigma}_{\mathbf{s}_l,\mathbf{u}_{\mathcal{L}_{l-1} }} = \boldsymbol{\delta}_l\mathbf{\Sigma}_{\mathbf{s}}\bar{\mathbf{I}}_l^T, \;
 \mathbf{\Sigma}_{\mathbf{u}_{\mathcal{L}_{l-1} }} = \bar{\mathbf{I}}_l\mathbf{\Sigma}_{\mathbf{u}_{\mathcal{L}}}\bar{\mathbf{I}}_l^T$,
$\mathbf{\Sigma}_{\mathbf{z}}=\mathbf{W}\mathbf{\Sigma}_{\mathbf{s}}\mathbf{W}^H$,
$ \mathbf{\Sigma}_{\mathbf{z},\mathbf{u}_{\mathcal{L}}}=\mathbf{W}\mathbf{\Sigma}_{\mathbf{s}}$,
$\mathbf{\Sigma}_{\mathbf{u}_{\mathcal{L}}}=\mathbf{\Sigma}_{\mathbf{s}} + \text{diag}[\mathbf{K}_{\mathcal{L}}]$.
\end{theorem}

\noindent \textbf{Proof:} 
We evaluate Theorem \ref{th:InnoRout} by considering jointly Gaussian sources and test channels $(\mathbf{S}_1,\ldots\mathbf{S}_L,\mathbf{U}_1,\ldots,\mathbf{U}_L)$ satisfying $p(\mathbf{s}_1,\ldots, \mathbf{s}_L)\prod_{l=1}^{L} p(\mathbf{u}_l|\mathbf{s}_l,\mathbf{u}_1,\ldots,\mathbf{u}_{l-1})$ and the minimum mean square error (MMSE) estimator $\mathbf{\hat{Z}} = \mathrm{E}[\mathbf{Z}|\mathbf{U}_{\mathcal{L}}]$ as reconstruction function $g$, where we define $\mathbf{U}_{\mathcal{L}_{l-1}}\triangleq [\mathbf{U}_1,\ldots,\mathbf{U}_{l-1}]$ and $\dv U_{\mc L}\triangleq \dv U_{\mc L_L}$. Note that MMSE reconstruction is optimal under \eqref{eq:sumdist}, while considering jointly Gaussian test channels might be suboptimal in general. 
First we derive a lower bound on the achievable distortion. We have
\begin{align}
B_l&\triangleq  I(\textbf{S}_{l};\textbf{U}_{l}|\mathbf{U}_{\mathcal{L}_{l-1}} )\\
& = I(\textbf{J}_{l};\textbf{U}_{l}|\mathbf{U}_{\mathcal{L}_{l-1}} )\label{eq:Innovation}\\
& = I(\textbf{J}_{l}; \mathbf{U}_{\mathcal{L}_{l-1}},\textbf{U}_{l})\label{eq:Orth}\\
&\geq I(\textbf{J}_{l}; \textbf{U}_{l}),\label{eq:innoineq}
\end{align}
where  in \eqref{eq:Innovation} we define the MMSE error $\mathbf{J}_l\triangleq \textbf{S}_{l}-\mathrm{E}[\mathbf{S}_l|\mathbf{U}_{\mathcal{L}_{l-1}}]$, which is Gaussian distributed  as $\mathbf{J}_l\sim \mathcal{CN}(\mathbf{0},\mathbf{\Sigma}_{\mathbf{s}_l|\mathbf{u}_{\mathcal{L}_{l-1} }})$; \eqref{eq:Orth} follows due to the orthogonality principle \cite{elGamal:book}, and due to the fact that for Gaussian random variables, orthogonality implies independence of $\mathbf{J}_l$ and $\mathbf{U}_{\mathcal{L}_{l-1}}$.

For the fixed test channels, let us choose matrix $\mathbf{K}_l\succeq \mathbf{0}$  such that for $l=1,\ldots, L$ 
\begin{align}
\text{cov}(\mathbf{J}_l|\mathbf{U}_l) = \mathbf{\Sigma}_{\mathbf{s}_l|\mathbf{u}_{\mathcal{L}_{l-1} }}^{1/2} 
\mathbf{K}_l^{1/2}(\mathbf{\Sigma}_{\mathbf{s}_l|\mathbf{u}_{\mathcal{L}_{l-1} }}+\mathbf{K}_l)^{-1}\mathbf{K}_l^{1/2}
\mathbf{\Sigma}_{\mathbf{s}_l|\mathbf{u}_{\mathcal{L}_{l-1} }}^{1/2}. \label{eq:matrixeq}
\end{align}
Note that such $\mathbf{K}_l$ always exists since $\mathbf{0} \preceq \text{cov}(\mathbf{J}_l|\mathbf{U}_l)\preceq\mathbf{\Sigma}_{\mathbf{s}_l|\mathbf{u}_{\mathcal{L}_{l-1} }} $, and can be found explicitly as follows.  After some straightforward algebraic manipulations, \eqref{eq:matrixeq} can be written as $\mathbf{K}_l^{1/2}\mathbf{A}\mathbf{K}_l^{1/2,H} = \mathbf{\Sigma}_{\mathbf{s}_l|\mathbf{u}_{\mathcal{L}_{l-1} }}$, where $\mathbf{A}\triangleq \mathbf{\Sigma}_{\mathbf{s}_l|\mathbf{u}_{\mathcal{L}_{l-1} }}^{1/2} \text{cov}^{-1}(\mathbf{J}_l|\mathbf{U}_l) \mathbf{\Sigma}_{\mathbf{s}_l|\mathbf{u}_{\mathcal{L}_{l-1} }}^{1/2} -\mathbf{I}$. Note that $\mathbf{A}\succeq \mathbf{0}$, and let $\mathbf{A}=\mathbf{V}\boldsymbol\Lambda_A \mathbf{V}^H$ and $\mathbf{\Sigma}_{\mathbf{s}_l|\mathbf{u}_{\mathcal{L}_{l-1} }} = \mathbf{V}'\boldsymbol\Lambda_{\Sigma_l}\mathbf{V}'^H$. 
 Then, it follows that $\mathbf{K}_l$ is given by $\mathbf{K}_l=\mathbf{K}_l^{1/2}\mathbf{K}_l^{1/2,H}$, where 
$\mathbf{K}_l^{1/2}=\mathbf{V}'\boldsymbol\Lambda_K^{1/2}\mathbf{V}^{H}$ and $\boldsymbol\Lambda_K = \boldsymbol\Lambda_{\Sigma_l}/\boldsymbol\Lambda_A $.
Then, from \eqref{eq:innoineq}, we have
\vspace{-2mm}
\begin{align}
B_l&\geq h(\mathbf{J}_l)-h(\mathbf{J}_l|\mathbf{U}_l)\\
&\geq \log|\mathbf{\Sigma}_{\mathbf{s}_l|\mathbf{u}_{\mathcal{L}_{l-1} }}|- \log| \text{cov}(\mathbf{J}_l|\mathbf{U}_l)|\\
&= \log|\mathbf{\Sigma}_{\mathbf{s}_l|\mathbf{u}_{\mathcal{L}_{l-1} }}+\mathbf{K}_l| - \log|\mathbf{K}_l|.\label{eq:K_eq}
\end{align}
\vspace{-2mm}
The distortion is lower bounded as
\vspace{-2mm}
\begin{align}
D
 \geq \mathrm{Tr}\{\mathrm{E}[(\mathbf{Z}-\mathrm{E}[\mathbf{Z}|\mathbf{U}_{\mathcal{L}}])(\mathbf{Z}-\mathrm{E}[\mathbf{Z}|\mathbf{U}_{\mathcal{L}}])^H]\}
 &=\mathrm{Tr}\{\mathbf{\Sigma}_{\mathbf{z}}-\mathbf{\Sigma}_{\mathbf{z},\mathbf{u}_{\mathcal{L}}}\mathbf{\Sigma}_{\mathbf{u}_{\mathcal{L} }}^{-1}\mathbf{\Sigma}_{\mathbf{z},\mathbf{u}_{\mathcal{L}}}^H \},\label{eq:MMSEGauss}
\end{align}
where \eqref{eq:MMSEGauss} follows due to the linearity of the MMSE estimator for jointly Gaussian variables.

The lower bound given by \eqref{eq:K_eq} and \eqref{eq:MMSEGauss} is achievable by letting $\mathbf{U}_l = \mathbf{S}_l+\mathbf{Q}_l$,  with $\mathbf{Q}_l\sim\mathcal{CN}(\mathbf{0},\mathbf{K}_l)$, and independent of all other variables, as follows
\vspace{-4mm}
\begin{align}
B_l&= I(\textbf{S}_{l};\textbf{U}_{l}|\mathbf{U}_{\mathcal{L}_{l-1}} )\\
 &= h(\textbf{U}_{l}|\mathbf{U}_{\mathcal{L}_{l-1}})-h(\textbf{U}_{l}|\mathbf{U}_{\mathcal{L}_{l-1}},\textbf{S}_{l} )\\
&= h(\textbf{S}_{l}-\mathrm{E}[\mathbf{S}_l|\mathbf{U}_{\mathcal{L}_{l-1}}]+\mathbf{Q}_l|\mathbf{U}_{\mathcal{L}_{l-1}} )-h(\textbf{Q}_{l})\label{eq:testGaussian}\\
&= h(\textbf{S}_{l}-\mathrm{E}[\mathbf{S}_l|\mathbf{U}_{\mathcal{L}_{l-1}}] +\mathbf{Q}_l)-h(\textbf{Q}_{l})\label{eq:MMSEOrth}\\
&= \log|\mathbf{\Sigma}_{\mathbf{s}_l|\mathbf{u}_{\mathcal{L}_{l-1} }}+\mathbf{K}_l| - \log|\mathbf{K}_l|,\label{eq:MMSEOrthGauss}
\end{align}
where \eqref{eq:testGaussian} follows since $\mathbf{U}_l = \mathbf{S}_l+\mathbf{Q}_l$ and \eqref{eq:MMSEOrth} is due to the orthogonality principle.
In the case $\text{cov}(\mathbf{J}_l|\mathbf{U}_l)=\mathbf{\Sigma}_{\mathbf{s}_l|\mathbf{u}_{\mathcal{L}_{l-1} }} $, we have $B_l=0$ which can be trivially achieved by letting $\mathbf{U}_l=\emptyset$.
 
Optimizing over the positive semidefinite covariance matrices $\mathbf{K}_1,\ldots,\mathbf{K}_{L}\succeq 
\mathbf{0}$ gives the desired minimum distortion $D$ in Theorem \ref{lem:InnoGaus}. This completes the proof.
\qed

\begin{figure}[!t]
\begin{minipage}[t]{0.5\linewidth}
    \centering
\includegraphics[width=0.995\textwidth]{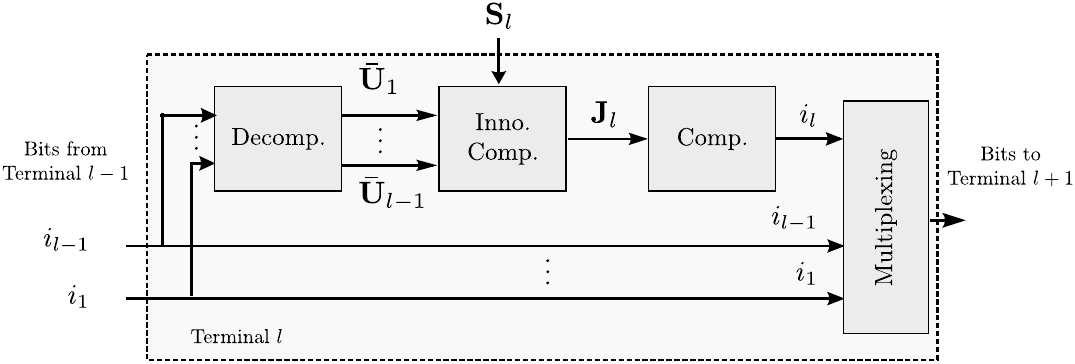}
\vspace{-5mm}	
\caption{Improved Routing scheme for C-MIMO.}\label{fig:Inno}
\end{minipage}
\hspace{0.1cm}
\begin{minipage}[t]{0.5\linewidth} 
    \centering
	\includegraphics[width=0.995\textwidth]{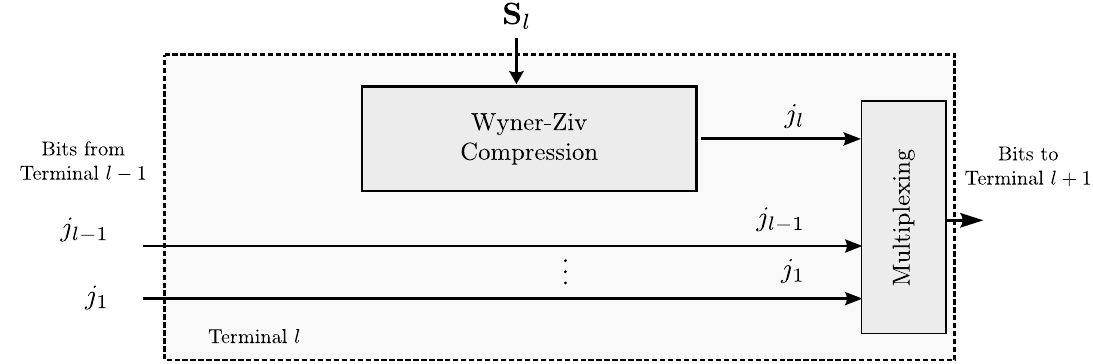}
	\vspace{-5mm}	
\caption{Wyner-Ziv Routing scheme for C-MIMO.}\label{fig:WZR}
\end{minipage}        
\end{figure}

The IR scheme in Section \ref{ssec:InnoRouting} requires joint compression at RRU $l$ of the observed source $\mathbf{S}_l$ and the previous compression codewords $\mathbf{U}_1,\ldots, \mathbf{U}_{l-1}$ to generate the compression codeword $\mathbf{U}_l$. However, for the Gaussian C-MIMO, it is shown next that the sum-distortion $D_{\mathrm{IR}}$ in Theorem \ref{lem:InnoGaus} can also be achieved by applying at each RRU separate decompression of the previous compression codewords, the innovation sequence computation $\mathbf{J}_l$, followed by independent compression of $\mathbf{J}_l$ into a codeword $\mathbf{\bar{U}}_{l}$, which is independent of the previous compression codewords $\mathbf{\bar{U}}_{1},\ldots, \mathbf{\bar{U}}_{l-1}$, as follows. See Figure \ref{fig:Inno}. At RRU $l$:
\begin{itemize}
\item Upon receiving bits $m_{l-1}$, decompress $\mathbf{\bar{U}}_1,\ldots,\mathbf{\bar{U}}_{l-1}$. 
\item Compute the innovation sequence $\mathbf{J}_l\triangleq \mathbf{S}_l-\mathrm{E}[\mathbf{S}_l|\mathbf{\bar{U}}_1,\ldots,\mathbf{\bar{U}}_{l-1}]$. 
\item Compress $\mathbf{J}_l$ at $B_l$ bits per sample independently of $\mathbf{\bar{U}}_1,\ldots, \mathbf{\bar{U}}_{l-1}$ using a codeword $\mathbf{\bar{U}}_l$, where  $\mathbf{\bar{U}}_l=\mathbf{J}_l+\mathbf{Q}_l$, with $\mathbf{Q}_l\sim\mathcal{CN}(\mathbf{0},\mathbf{K}_l)$ independent of each other.
\end{itemize}

Note that $\mathbf{J}_l$ corresponds to the MMSE error of estimating $\mathbf{S}_l$ from $(\mathbf{\bar{U}}_1,\ldots, \mathbf{\bar{U}}_{l-1})$, and is an i.i.d. zero-mean Gaussian sequence distributed as $\mathbf{J}_l\sim\mathcal{CN}(\mathbf{0},\mathbf{\Sigma}_{\mathbf{j},l})$.
\begin{proposition}
For the Gaussian C-MIMO model, separate decompression, innovation computation and independent innovation compression achieves the minimum distortion $D_{\mathrm{IR}}(R_1,\ldots, R_L)$ characterized by the distortion-rate function in Theorem  \ref{lem:InnoGaus}.
\end{proposition}
\noindent \textbf{Proof:} 
We show that any distortion $D$ achievable for a pmf $p(\mathbf{s}_1,\ldots, \mathbf{s}_L)\prod_{l=1}^{L} p(\mathbf{u}_l|\mathbf{s}_l,\mathbf{u}_1,\ldots,\mathbf{u}_{l-1})$ and the corresponding $(B_1,\ldots, B_L)$   in Theorem \ref{lem:InnoGaus} is also achievable  with separate decompression, innovation computation and compression as detailed above.   From standard arguments, compressing $\mathbf{J}_l$ at $B_l$ bits requires
\begin{align}
B_{l}&\geq I(\mathbf{J}_l;\mathbf{\bar{U}}_l)\\
& = h(\mathbf{\bar{U}}_l)-h(\mathbf{\bar{U}}_l|\mathbf{J}_l)\\
&= \log|\mathbf{\Sigma}_{\mathbf{s}_l|\mathbf{u}_{\mathcal{L}_{l-1} }} +\mathbf{K}_l| - \log|\mathbf{K}_l|,\label{eq:EqSigma}
\end{align}
where \eqref{eq:EqSigma} follows since $\mathbf{\Sigma}_{\mathbf{j},l}=\mathbf{\Sigma}_{\mathbf{s}_l|\mathbf{u}_{\mathcal{L}_{l-1} }} $, which follows since RRU $l$ can compute $\mathbf{U}_{l'} = \bar{\mathbf{U}}_{l'} + \mathrm{E}[\mathbf{S}_{l'}|\mathbf{U}_1, \ldots, \mathbf{U}_{l'-1}]$ for  $l'=1,\ldots, l$, which is distributed  as the test channels  $\mathbf{U}_{l'}=\mathbf{S}_{l'}+\mathbf{Q}_{l'}$ and thus $\mathrm{E}[\mathbf{S}_l|\mathbf{U}_{\mathcal{L}_{l-1}}] = \mathrm{E}[\mathbf{S}_l|\mathbf{\bar{U}}_{\mathcal{L}_{l-1}}]$. 

The distortion between $\mathbf{Z}$ and its estimation from $\mathbf{\bar{U}}_{\mathcal{L}}$ satisfies
\begin{align}
D&\geq\mathrm{Tr}\{\mathrm{E}[(\mathbf{Z}-\mathrm{E}[\mathbf{Z}|\mathbf{\bar{U}}_{\mathcal{L}}])(\mathbf{Z}-\mathrm{E}[\mathbf{Z}|\mathbf{\bar{U}}_{\mathcal{L}}])^H]\}\\
	&=\mathrm{Tr}\{\mathrm{E}[(\mathbf{Z}-\mathrm{E}[\mathbf{Z}|\mathbf{U}_{\mathcal{L}}])(\mathbf{Z}-\mathrm{E}[\mathbf{Z}|\mathbf{U}_{\mathcal{L}}])^H]\}\\
 &=\mathrm{Tr}\{\mathbf{\Sigma}_{\mathbf{z}}-\mathbf{\Sigma}_{\mathbf{z},\mathbf{u}_{\mathcal{L}}}\mathbf{\Sigma}_{\mathbf{u}_{\mathcal{L} }}^{-1}\mathbf{\Sigma}_{\mathbf{z},\mathbf{u}_{\mathcal{L}}}^H \}.
\end{align}
Thus, any achievable distortion $D$ for given $p(\mathbf{s}_1,\ldots, \mathbf{s}_L)\prod_{l=1}^{L} p(\mathbf{u}_l|\mathbf{s}_l,\mathbf{u}_1,\ldots,\mathbf{u}_{l-1})$ and fixed $(B_1,\ldots, B_L)$ in Theorem \ref{lem:InnoGaus} is achievable by separate decompression, innovation computation and  independent compression of the innovation. This completes the proof.
\qed

Determining the optimal covariance matrices $(\mathbf{K}_1,\ldots,\mathbf{K}_L)$ achieving $D_{\mathrm{IR}}(R_1,\ldots,R_L)$ in Theorem \ref{lem:InnoGaus} requires a joint optimization, which is generally not simple. Next, we propose a method to successively obtain a feasible solution $(\mathbf{K}^*_1,\ldots,\mathbf{K}^*_L)$ and the corresponding minimum distortion $D_{\mathrm{IR-S}}^{*}(R_1,\ldots,R_L)$ for given $(R_1,\ldots, R_L)$:
\begin{enumerate}
\item For a given fronthaul tuple $(R_1,\dots, R_L)$, fix non-negative  $B_1,\ldots, B_L$, satisfying $R_l\geq B_1+\ldots+B_l$, for  $l=1,\ldots, L$.
\item For such $(B_1,\ldots,B_L)$, sequentially find $\mathbf{K}^*_l$ from RRU $1$ to RRU $L$ as the $\mathbf{K}_l$ minimizing the distortion between the innovation $\mathbf{J}_l$ and its reconstruction  as follows. At RRU $l$, for given $\mathbf{K}^*_1,\ldots,\mathbf{K}_{l-1}^*$ and  $B_l$, $\mathbf{K}_l^*$ is found from the covariance matrix $\mathbf{K}_l$ minimizing 
\vspace{-3.5mm}
\begin{align}\label{eq:RDProblem}
D_l (B_l) \triangleq  &\min_{\mathbf{K}_l}\mathrm{Tr}\{\mathrm{E}[(\mathbf{J}_l-\mathrm{E}[\mathbf{J}_l|\mathbf{\bar{U}}_l])(\mathbf{J}_l-\mathrm{E}[\mathbf{J}_l|\mathbf{\bar{U}}_l])^H]\}
\\ &\text{ s.t. }B_l\geq  \log|\mathbf{\Sigma}_{\mathbf{s}_l|\mathbf{u}_{\mathcal{L}_{l-1} }}+\mathbf{K}_l|/|\mathbf{K}_l|.\nonumber
\end{align}
Note that \eqref{eq:RDProblem} corresponds to the distortion-rate problem of compressing a Gaussian vector source $\mathbf{J}_l\sim \mathcal{CN}(\mathbf{0},\boldsymbol{\Sigma}_{\mathbf{j},l})$ at $B_l$ bits and its solution is given below in Proposition \ref{prop:RDInnov}.

\item Compute the achievable distortion $D_{\mathrm{IF}}^{\mathrm{S}}(B_1,\ldots, B_L)$ by evaluating $\mathrm{Tr}\{\mathbf{\Sigma}_{\mathbf{z}}-\mathbf{\Sigma}_{\mathbf{z},\mathbf{u}_{\mathcal{L}}}\mathbf{\Sigma}_{\mathbf{u}_{\mathcal{L} }}^{-1}\mathbf{\Sigma}_{\mathbf{z},\mathbf{u}_{\mathcal{L}}}^H \}$ as in Theorem \ref{th:InnoRout} with the chosen covariance matrices $(\mathbf{K}_1^*,\ldots, \mathbf{K}_L^*)$.
\item Compute $D_{\mathrm{IR-S}}^{*}(R_1,\ldots,R_L)$ as the minimum $D_{\mathrm{IF}}^{\mathrm{S}}(B_1,\ldots, B_L)$ over $(B_1,\ldots, B_L)$ satisfying the fronthaul constraints $R_l\geq B_1+\cdots + B_l$ for $l=1,\ldots, L$ .
\end{enumerate}

 The solution for the distortion-rate problem in \eqref{eq:RDProblem} is standard and given next for completeness.
\begin{proposition}\label{prop:RDInnov}
Given $\mathbf{K}^*_1,\ldots,\mathbf{K}_{l-1}^*$, let  $\mathbf{\Sigma}_{\mathbf{j},l}=\mathbf{\Sigma}_{\mathbf{s}_l|\mathbf{u}_{\mathcal{L}_{l-1} }} =\mathbf{V}_l\boldsymbol\Lambda_{J}\mathbf{V}_l^H$, where $\mathbf{V}_l^H\mathbf{V}_l = \mathbf{I}$ and  $\mathbf{\Lambda}_J\triangleq\text{diag}[\lambda_1^J,\ldots,\lambda_K^J]$.
The optimal distortion \eqref{eq:RDProblem} is $D_l = \sum_{k=1}^{K}\min\{\lambda,\lambda_k^J\}$ where
  $\lambda>0$ is  the solution to
\begin{align}\label{eq:lambdaeq}
B_l = \sum_{k=1}^{K}\log^+\left(\frac{\lambda^{J}_k}{\lambda}\right),
\end{align}
and is achieved with $\mathbf{K}_l^*= \mathbf{V}_l\mathbf{\Lambda}\mathbf{V}_l^H$, where $\mathbf{\Lambda}=\text{diag} [\lambda_1^Q,\ldots,\lambda_K^Q]$ and $\lambda_k^Q=\min\{\lambda,\lambda_k^J\}/(\lambda_k^J-\min\{\lambda,\lambda_k^J\})$\footnote{Note the slight abuse of notation. If for the $k$-th uncorrelated components we have $\lambda\leq \lambda_{k}^J$,  in the achievability we have $\lambda_k^Q=\infty$. It should be understood that the $k$-th component is not assigned any bit for compression. This is in line with \eqref{eq:lambdaeq} as the number of bits assigned for the $k$-th component is given by $\log^+\left(\lambda^{J}_k/\lambda\right)=0$.}. 
\end{proposition}
\noindent \textbf{Outline Proof:} 
The minimization of the RD problem in \eqref{eq:RDProblem} is standard, e.g. \cite{Cover:book},  and well known to be achieved by uncorrelating the vector source $\mathbf{J}_l$ into $K$ uncorrelated components as $\mathbf{J}_l'= \mathbf{V}^{H}\mathbf{J}_l \sim \mathcal{CN}(\mathbf{0},\mathbf{\Lambda}_J)$. Then, the available $B_l$ bits are distributed over the $K$ parallel source components $\mathbf{J}_l'$ by solving the reverse water-filling problem
\begin{align}
D_l = &\min_{d_1,\ldots, d_K\geq 0}\sum_{k=1}^{K}d_k  &\text{ s.t. }B_l=  \sum_{k=1}^{K}\log^+\left(\frac{\lambda^{J}_k}{d_k}\right),\nonumber
\end{align}
 The solution to this problem is given by $d_k = \min\{\lambda^{J}_k,\lambda\}$, where $\lambda>0$ satisfies \eqref{eq:lambdaeq}. The optimality of $\mathbf{K}_l^*$ follows since $D_l$ is achieved with $\mathbf{K}_l^*$ as stated in Proposition~\ref{prop:RDInnov} \cite{Cover:book,tian2009remote}.
\qed


\subsection{Centralized Beamforming with Successive Wyner-Ziv}

In this section, we consider the distortion-rate function of the WZR scheme in Corollary~\ref{th:WZR} for centralized beamforming. Similarly to IR, each RRU forwards a compressed version of its observation to the CP, which  estimates the receive-beamforming signal $\mathbf{Z}$ from the decompressed observations. 
Next theorem shows that WZR achieves the same distortion-rate function as the IR scheme under jointly Gaussian test channels.
\begin{theorem}\label{lem:BTGaus}
The distortion-rate function of the WZR scheme $D_{\mathrm{WZR}}(R_1,\ldots,R_L)$ with jointly Gaussian test channels, is the same as the distortion-rate function of the IR scheme with Gaussian test channels in Theorem 
\ref{lem:InnoGaus}, i.e., $D_{\mathrm{WZR}}(R_1,\ldots,R_L) = D_{\mathrm{IR}}(R_1,\ldots,R_L)$
\end{theorem}

\noindent \textbf{Outline Proof:} 
Since $\mathcal{R}_{\mathrm{WZR}}(D)\subseteq \mathcal{R}_{\mathrm{IR}}(D)$, we only need to show that any distortion $D$ achievable with IR in Theorem \ref{lem:InnoGaus} is also achievable with WZR. For fixed $(B_1,\ldots, B_L)$ and $p(\mathbf{s}_1,\ldots, \mathbf{s}_L)\prod_{l=1}^{L} p(\mathbf{u}_l|\mathbf{s}_l,\mathbf{u}_1,\ldots,\mathbf{u}_{l-1})$ with IR in Theorem \ref{lem:InnoGaus},  the minimum distortion is achieved by considering a test channel $\mathbf{U}_l = \mathbf{S}_l+\mathbf{Q}_l$. Since this test channel is also in the class of test channels $\prod_{l=1}^{L} p(\mathbf{u}_l|\mathbf{s}_l)$ of WZR, it follows that any achievable distortion $D$ for fixed $(B_1,\ldots, B_L)$ and $p(\mathbf{s}_1,\ldots, \mathbf{s}_L)\prod_{l=1}^{L} p(\mathbf{u}_l|\mathbf{s}_l,\mathbf{u}_1,\ldots,\mathbf{u}_{l-1})$ in Theorem \ref{lem:InnoGaus} is achievable with WZR.
\qed

\subsection{In-Network Processing for Distributed Beamforming}\label{ssec:InlineBeam}

In this section, we study the distortion-rate function  of the IP scheme in Section \ref{ssec:InlineProcessing} for distributed beamforming. At each RRU, the received signal from the previous terminal is jointly compressed with the observation and forwarded to the next RRU.  While the optimal joint compression per RRU along the cascade remains an open problem, even for independent observations \cite{elGamal:book}, we propose to  gradually compute the desired function $\mathbf{Z}$ by reconstructing at each RRU  parts of $\mathbf{Z}$. In particular, compression at RRU $l-1$ is designed such that RRU $l$ reconstructs from $\mathbf{S}_l$ and the received bits an estimate of the part of the function:
\begin{align}\label{eq:partial_beamf}
\mathbf{Z}_l\triangleq \mathbf{W}_1 \mathbf{S}_1+\cdots+\mathbf{W}_l\mathbf{S}_{l}.
\end{align}

The design of the  compression is done successively. Assuming $\mathbf{U}^*_{\mathcal{L}_{l-1}}\triangleq (\mathbf{U}_{1}^*,\ldots, \mathbf{U}_{l-1}^*)$ are fixed, at RRU $l$, $\mathbf{U}_l^*$, is obtained as the solution to the following distortion-rate problem:
\begin{align}
D_{l}(R_l) =& \min_{p(\mathbf{u}_l|\mathbf{s}_l,\mathbf{u}^*_{l-1})}\mathrm{Tr}\{\mathrm{E}[(\mathbf{Z}_{l}-\mathbf{\hat{Z}}_{l})(\mathbf{Z}_l-\mathbf{\hat{Z}}_{l})^H]\}\label{eq:RD_remote_function}\\
&\text{s.t. }R_l\geq  I(\mathbf{S}_l,\mathbf{U}_{l-1}^*;\mathbf{U}_l|\mathbf{S}_{l+1})\label{eq:RD_remote_function_2}.
\end{align}
Problem \eqref{eq:RD_remote_function}-\eqref{eq:RD_remote_function_2} corresponds to the distortion-rate function of the Wyner-Ziv type source coding problem of lossy reconstruction of function $\mathbf{Z}_l$ as $\mathbf{\hat{Z}}_l$, which is a function of the encoder observation $\mathbf{S}_l,\mathbf{U}_{l-1}^*$ when side information $\mathbf{S}_{l+1}$, is  available at the decoder \cite{Yamamoto:IT:1982}. Proposition \ref{th:RemoteProblem_inno} given below characterizes the optimal test channel at RRU $l$ given $\mathbf{U}^*_{\mathcal{L}_{l-1}}$, i.e.,  $\mathbf{U}_{l}^*$, and shows that it is Gaussian distributed as,
\vspace{-2mm}
\begin{align}
\mathbf{U}^{*}_l =\mathbf{P}_l[\mathbf{U}^*_{l-1};\mathbf{W}_l\mathbf{S}_l]^H + \mathbf{Q}_l,\label{eq:optimaltest_Inline}
\end{align}
\vspace{-2mm}
where $\mathbf{P}_l=[\mathbf{P}_l^{U},\mathbf{P}_{l}^{S}]$ and $\mathbf{Q}_l\sim\mathcal{CN}(\mathbf{0},\mathbf{K}_l)$. 
\begin{remark}
Operationally,  \eqref{eq:optimaltest_Inline} indicates that the optimal codeword $\mathbf{U}_l^*$ at RRU $l$ can be obtained by compressing a linear combination of the decompressed signal $\mathbf{U}_{l-1}^*$ and a beamformed version of the observation $\mathbf{W}_l\mathbf{S}_l$ with the linear combination $\mathbf{P}_l$. This is exploited below in Proposition~\ref{Prop:IPProp}.
\end{remark}

On the other hand, let $\mathbf{\Pi}^{U}_{l,l'} = \mathbf{P}_{l}^{U}\cdots \mathbf{P}_{l'}^{U}$ and the quantization noises $\mathbf{\bar{Q}} = [\mathbf{Q}_1^T,\ldots,\mathbf{Q}_{L}^T]^{T}$,
\begin{align}
\mathbf{\bar{P}}_l^{S} &\triangleq 
\left[ \mathbf{\Pi}_{l,2}^{U}\mathbf{P}_1^S\mathbf{W}_1, \ldots,
  \mathbf{\Pi}_{l,l}^{U}\mathbf{P}_{l-1}^S\mathbf{W}_{l-1}, \mathbf{P}_{l}^{S}\mathbf{W}_l,\mathbf{0},\ldots,\mathbf{0} \right],\nonumber\\
\mathbf{\bar{P}}_l^Q &\triangleq 
\left[ \mathbf{\Pi}_{l,2}^{U}\mathbf{P}_1^S,\mathbf{\Pi}_{l,3}^{U}\mathbf{P}_2^{S}, \ldots,
  \mathbf{P}_{l}^{U}\mathbf{P}_{l-1}^S,\mathbf{0}, \mathbf{0},\ldots,\mathbf{0} \right].
\end{align}
Then,  we can write $\mathbf{U}_l^*$ in   \eqref{eq:optimaltest_Inline} as $\mathbf{U}_l^* = \mathbf{R}_l+\mathbf{Q}_l$ where
\begin{align}
\mathbf{R}_l &\triangleq [\mathbf{P}_l^{U},\mathbf{P}_{l}^{S}][\mathbf{U}_{l-1}^*;\mathbf{W}_l\mathbf{S}_l]^H =\mathbf{\bar{P}}_l^S\mathbf{S}+\mathbf{\bar{P}}_l^Q\bar{\mathbf{Q}}_l.\label{eq:remotesource}
\end{align}
\begin{remark}
Equation \eqref{eq:remotesource} highlights that due to the successive decompression and recompression performed at each RRU, the quantization noises $\mathbf{Q}_l$ propagate throughout the cascade. The linear combination of the locally beamformed signal $\tilde{\mathbf{S}}_l=\mathbf{W}_l\mathbf{S}_l$ and the decompressed signal $\mathbf{U}_{l-1}^*$ can be seen as a noisy observation of the remote sources $\mathbf{S}$, through an additive channel with channel coefficients $\mathbf{\bar{P}}_l^{S}$ and correlated noise $\mathbf{\bar{P}}_l^Q\bar{\mathbf{Q}}_l$. 
This noisy signal is used as an estimate of the partial beamformed signal \eqref{eq:partial_beamf} to be reconstructed at the next RRU. 
\end{remark}

Next proposition characterizes the optimal test channel $\mathbf{U}_l^*$ in \eqref{eq:optimaltest_Inline} with $\mathbf{P}_{l}^*$  and  $\mathbf{K}_{l}^*$ for given test channels $\mathbf{U}^*_{\mathcal{L}_{l-1}}$ with their corresponding $\mathbf{P}_{1}^*,\ldots, \mathbf{P}_{l-1}^*$ and $\mathbf{K}_1^*, \ldots , \mathbf{K}_{l-1}^*$.

\begin{proposition}\label{th:RemoteProblem_inno}
Let $\mathbf{F}_l \triangleq [\mathbf{U}_{l-1}^*;\mathbf{S}_l]$, $\mathbf{T}_{\mathbf{z}_l,\mathbf{f}_l}\mathbf{\Sigma}_{\mathbf{f}_l|\mathbf{s}_{l+1}}\mathbf{T}_{\mathbf{\bar{z}}_l,\mathbf{f}_l}^H = \mathbf{V}_l\boldsymbol\Lambda_l^{D}\mathbf{V}_l^H$, where $\mathbf{\Lambda}^D=\text{diag}[\lambda_1^D,\ldots,\lambda_M^D]$ and $\mathbf{T}_{\mathbf{\bar{z}}_l,\mathbf{f}_l} = \mathbf{\Sigma}_{\mathbf{\bar{z}}_l,\mathbf{f}_l}\mathbf{\Sigma}_{\mathbf{f}_l}^{-1}$, and $ \mathbf{\Sigma}_{\mathbf{\bar{z}}_l,\mathbf{f}_l} = \mathbf{\bar{W}}_l\mathbf{\Sigma}_{\mathbf{s}}\mathbf{\bar{P}}^{S,H}_{l}$, where $\mathbf{\bar{W}}_l\triangleq[\mathbf{W}_1,\ldots, \mathbf{W}_l,\mathbf{0},\ldots,\mathbf{0}]$. The minimum distortion \eqref{eq:RD_remote_function} is $D_l(R_l) = \sum_{m=1}^{M}\min\{\lambda,\lambda_m^D\} + \mathrm{Tr}\{\mathbf{\Sigma}_{\mathbf{\bar{\mathbf{z}}_l}|\mathbf{f}_l\mathbf{s}_{l+1}}\}$ 
and $\lambda>0$ satisfies
\vspace{-1mm}
\begin{align}
R_l = \sum_{m=1}^{M}\log^+\left(\frac{\lambda^{D}_m}{\lambda}\right), \label{eq:InProcConst}
\end{align}
where
$\mathbf{\Sigma}_{\mathbf{\bar{\mathbf{z}}_l}|\mathbf{f}_l\mathbf{s}_{l+1}} = 
\mathbf{\Sigma}_{\mathbf{\bar{\mathbf{z}}_l}} - 
\mathbf{\Sigma}_{\mathbf{\bar{\mathbf{z}}_l},\mathbf{f}_l\mathbf{s}_{l+1}} 
\mathbf{\Sigma}_{\mathbf{\mathbf{f}_l\mathbf{s}_{l+1}}}^{-1} 
\mathbf{\Sigma}_{\mathbf{\bar{\mathbf{z}}_l},\mathbf{f}_l\mathbf{s}_{l+1}}^H  
$,
 $\mathbf{\Sigma}_{\mathbf{\bar{\mathbf{z}}_l},\mathbf{f}_l\mathbf{s}_{l+1}} = \mathbf{\bar{W}}_l\mathbf{\Sigma}_{\mathbf{s}}[\mathbf{\bar{P}}_{S}, \boldsymbol\delta_{l+1}]^H$,
$\mathbf{\Sigma}_{\mathbf{\hat{z}_l}\mathbf{s}_{l+1}} = [\mathbf{\Sigma}_{\mathbf{f}_l}, \mathbf{\bar{P}}_{l}^{S}\mathbf{\Sigma}_{\mathbf{s}}\boldsymbol\delta_{l+1}^H;
\boldsymbol\delta_{l+1}^H\mathbf{\Sigma}_{\mathbf{s}}\mathbf{\bar{P}}_{l}^{S,T},\mathbf{\Sigma}_{\mathbf{s}_{l+1}}]$, 
In addition, the minimum distortion in $D_l(R_l)$ is  achieved with $\mathbf{K}_l^*= \mathbf{V}_l\mathbf{\Lambda}^{Q}\mathbf{V}_l^H$ and $\mathbf{P}^*_l = \mathbf{T}_{\mathbf{\bar{z}}_l,\hat{\mathbf{z}}_l}$, where $\mathbf{\Lambda}^{Q}=\text{diag} [\lambda_1^Q,\ldots,\lambda_M^Q]$ is a diagonal matrix, with the $m$-th diagonal element $\lambda_m^Q=\min\{\lambda,\lambda_m^D\}/(\lambda_m^D-\min\{\lambda,\lambda_m^D\})$.
\end{proposition}
\noindent \textbf{Outline Proof:} 
The proof is similar to that of the remote Wyner-Ziv source coding problem for source reconstruction in \cite{tian2009remote}. We consider lossy function reconstruction. For  simplicity, we drop the RRU index $l$ in this proof and define $\mathbf{F}\triangleq [\mathbf{U}_{l-1}^{*},\mathbf{W}_l\mathbf{S}_l]$, $\mathbf{Y}\triangleq \mathbf{S}_{l+1}$, $\mathbf{\bar{Z}}\triangleq \mathbf{Z}_l$ and $\mathbf{U} = \mathbf{U}_l$.  

First, we obtain a lower bound on the achievable distortion. Let us define the MMSE filters
\vspace{-3mm}
\begin{align}
\mathbf{T}_{\mathbf{f},\mathbf{y}} &= \boldsymbol\Sigma_{\mathbf{f},\mathbf{y}}\boldsymbol\Sigma^{-1}_{\mathbf{y}},
\quad \text{and } \quad 
[\mathbf{T}_{\mathbf{\bar{z}},\mathbf{f}} \; \mathbf{T}_{\mathbf{\bar{z}},\mathbf{y} } ]= [\boldsymbol\Sigma_{\mathbf{\bar{z}},\mathbf{f}} \boldsymbol\Sigma_{\mathbf{\bar{z}},\mathbf{y}}  ]
\left[
\begin{array}{cc}
\boldsymbol\Sigma_{\mathbf{f}}&\boldsymbol\Sigma_{\mathbf{f},\mathbf{y}}\\
\boldsymbol\Sigma_{\mathbf{f},\mathbf{y}}^H & \boldsymbol\Sigma_{\mathbf{y}} 
\end{array}\right]^{-1}.
\end{align}
We have from the MMSE estimation of Gaussian vector sources \cite{elGamal:book},
\vspace{-2mm}
\begin{align}
\mathbf{F} &= \mathbf{T}_{\mathbf{f},\mathbf{y}}\mathbf{Y}+\mathbf{N}_1,\label{eq:MMSE_sideinfo}\\
\mathbf{\bar{Z}} &= \mathbf{T}_{\mathbf{\bar{z}},\mathbf{f}}\mathbf{F}+\mathbf{T}_{\mathbf{\bar{z}},\mathbf{y} }\mathbf{Y}+\mathbf{N}_2 = (\mathbf{T}_{\mathbf{\bar{z}},\mathbf{f}}\mathbf{T}_{\mathbf{f},\mathbf{y}}+ \mathbf{T}_{\mathbf{\bar{z}},\mathbf{y} }) \mathbf{Y}+\mathbf{T}_{\mathbf{\bar{z}},\mathbf{f}}\mathbf{N}_1+\mathbf{N}_2,\label{eq:MMSE_channel}
\end{align}
where $\mathbf{N}_1$ and $\mathbf{N}_2$ correspond to the MMSE errors and are zero-mean jointly Gaussian random vectors independent of each other, $\mathbf{N}_1$ is independent of $\mathbf{Y}$ and $\mathbf{N}_2$ is independent of $\mathbf{F}, \mathbf{Y}$ and have the covariance matrices given by
$\boldsymbol\Sigma_{\mathbf{N}_1}\triangleq \boldsymbol\Sigma_{\mathbf{f}|\mathbf{y}}$ and $\boldsymbol\Sigma_{\mathbf{N}_2}\triangleq \boldsymbol\Sigma_{\bar{\mathbf{z}}|\mathbf{f}, \mathbf{y}}$.

On the other distortion side, we have
\vspace{-3mm}
\begin{align}
D &\triangleq \mathrm{E}[(\mathbf{\bar{Z}}-\mathbf{\hat{Z}})(\mathbf{\bar{Z}}-\mathbf{\hat{Z}})^H]\\
&\geq \mathrm{E}[(\mathbf{\bar{Z}}-\mathrm{E}[\bar{\mathbf{Z}}|\mathbf{U},\mathbf{Y}])(\mathbf{\bar{Z}}-\mathrm{E}[\bar{\mathbf{Z}}|\mathbf{U},\mathbf{Y}])^H]\\
&=\textrm{Tr}\{\mathrm{E}[(\mathbf{T}_{\mathbf{\bar{z}},\mathbf{f}}\mathbf{N}_1 + \mathbf{N}_2-\hat{\mathbf{N}}_1)(\mathbf{T}_{\mathbf{\bar{z}},\mathbf{f}}\mathbf{N}_1 + \mathbf{N}_2-\mathbf{\hat{N}}_1)^H]\}\label{eq:Line1_1}\\
&=\textrm{Tr}\{\mathrm{E}[(\mathbf{T}_{\mathbf{\bar{z}},\mathbf{f}}\mathbf{N}_1-\hat{\mathbf{N}}_1)(\mathbf{T}_{\mathbf{\bar{z}},\mathbf{f}}\mathbf{N}_1-\mathbf{\hat{N}}_1)^H]\}+\mathrm{Tr}\{\boldsymbol\Sigma_{\bar{\mathbf{z}}|\mathbf{f}, \mathbf{y}}\}\label{eq:Line1_2}
\end{align}
where \eqref{eq:Line1_1} follows from \eqref{eq:MMSE_channel} and where we have defined $\hat{\mathbf{N}}_1\triangleq  \mathrm{E}[\bar{\mathbf{Z}}|\mathbf{U}]- (\mathbf{T}_{\mathbf{\bar{z}},\mathbf{f}}\mathbf{T}_{\mathbf{f},\mathbf{y}}+ \mathbf{T}_{\mathbf{\bar{z}},\mathbf{y} }) \mathbf{Y}$; \eqref{eq:Line1_2} follows from the independence of $\mathbf{N}_2$ from $\mathbf{N}_1$, $\mathbf{Y}$ and $\mathbf{F}$.

Next, let us define $\mathbf{N}' = \mathbf{V}^H\mathbf{T}_{\mathbf{\bar{z}},\mathbf{f}}\mathbf{N}_1$ and $\mathbf{\hat{N}}' \triangleq \mathbf{V}^H\mathbf{\hat{N}}_1$, where $\mathbf{V}$ follows from the eigenvalue decomposition
\vspace{-6mm}
\begin{align}
\mathbf{T}_{\mathbf{\bar{z}},\mathbf{f}}\boldsymbol\Sigma_{\mathbf{N}_1}\mathbf{T}_{\mathbf{\bar{z}},\mathbf{f}}^H = \mathbf{T}_{\mathbf{\bar{z}},\mathbf{f}}\mathbf{\Sigma}_{\mathbf{f}|\mathbf{y}}\mathbf{T}_{\mathbf{\bar{z}},\mathbf{f}}^H = \mathbf{V}\boldsymbol\Lambda^{D}\mathbf{V}^H.
\end{align}
 Note that $\mathbf{N}'$ has independent components of variance $\boldsymbol\Lambda^{\mathrm{D}}$.
Therefore,  from \eqref{eq:Line1_2} we have
\begin{align}
D&=\textrm{Tr}\{\mathrm{E}[(\mathbf{N}'-\hat{\mathbf{N}}')(\mathbf{N}'-\mathbf{\hat{N}}')^H]\}+\mathrm{Tr}\{\boldsymbol\Sigma_{\bar{\mathbf{z}}|\mathbf{f}, \mathbf{y}}\}\label{eq:eqOrtho}\\
& = \sum_{m=1}^{M}\mathrm{E}[(N_m'-\hat{N}_m')^2]+\mathrm{Tr}\{\boldsymbol\Sigma_{\bar{\mathbf{z}}|\mathbf{f}, \mathbf{y}}\},\label{eq:eqOrtho_2}
\end{align}
where \eqref{eq:eqOrtho} follows due to the orthonormality of $\mathbf{V}$.

On the other hand, we have
\vspace{-0mm}
\begin{align}
R_l&\geq I(\mathbf{U}^*,\mathbf{S};\mathbf{U}|\mathbf{Y})\\
&\geq I(\mathbf{V}^H\mathbf{T}_{\mathbf{\bar{z}},\mathbf{f}}\mathbf{F};\mathbf{U}|\mathbf{Y}) \label{eq:dataproc}\\
&=h(\mathbf{V}^H\mathbf{T}_{\mathbf{\bar{z}},\mathbf{f}}\mathbf{F}|\mathbf{Y})-h(\mathbf{V}^H\mathbf{T}_{\mathbf{\bar{z}},\mathbf{f}}\mathbf{F}|\mathbf{Y},\mathbf{U})\label{eq:Achievable}\\
&=h(\mathbf{V}^H\mathbf{T}_{\mathbf{\bar{z}},\mathbf{f}}\mathbf{N}_1)-h(\mathbf{V}^H\mathbf{T}_{\mathbf{\bar{z}},\mathbf{f}}\mathbf{N}_1|\mathbf{Y},\mathbf{U})\label{eq:conditional}\\
&=\sum_{m=1}^{K}h(N_m')-h(N_m'|\mathbf{Y},\mathbf{U},\mathbf{N}_{1,1}^{',m-1})\label{eq:definitionNprime}\\
&\geq \sum_{m=1}^{M}h(N_m')-h(N_m'|\mathbf{Y},\mathbf{U})\label{eq:entropy_reduce}\\
&= \sum_{m=1}^{M}I(N_m';\mathbf{Y},\mathbf{U})\\
&\geq  \sum_{m=1}^{M}I(N_m';\hat{N_m'})\label{eq:dataproc_2}
\end{align}
where \eqref{eq:dataproc} follows due to the data processing inequality; \eqref{eq:conditional} is due to \eqref{eq:MMSE_sideinfo} and the orthogonality principle of the MMSE estimator; \eqref{eq:definitionNprime} is due to the definition of $\mathbf{N}'= (N_1',\ldots, N_M')$; \eqref{eq:entropy_reduce} follows since conditioning reduces entropy; and \eqref{eq:dataproc_2} is due to the data processing inequality and since $\hat{N}_m$, where $\mathbf{\hat{N}} = (\hat{N}_1',\ldots, \hat{N}_M')$, is a function of $\mathbf{Y},\mathbf{U}$.

It follows from \eqref{eq:eqOrtho_2} and \eqref{eq:dataproc_2}  that $D_l$ is lower bounded by the sum-distortion $D$ of compressing  $N_m'\sim\mathcal{CN}(0,\lambda^D_m)$, $m=1,\ldots, M$, given as a  reverse water filling problem with a modified distortion 
$\tilde{D}\triangleq D-\mathrm{Tr}\{\boldsymbol\Sigma_{\bar{\mathbf{z}}|\mathbf{f}, \mathbf{y}}\}$,
so that $N_m$ is reconstructed with distortion $d_m\triangleq \mathrm{E}[(N_m'-\hat{N}_m')^2]$,
\vspace{-5mm}
\begin{align}
 \tilde{D}  = \min_{d_1,\ldots,d_M>0}\sum_{m=1}^{M}d_m  \quad \text{s.t. } R(D)=\sum_{m=1}^{M}\log\left(\frac{\lambda^{D}_m}{d_m}\right). 
\end{align}
Note that if $D<\mathrm{Tr}\{\boldsymbol\Sigma_{\bar{\mathbf{z}}|\mathbf{f}, \mathbf{y}}\}$, then $R(D)=\infty$ and if $D>\mathrm{Tr}\{\boldsymbol\Sigma_{\bar{\mathbf{z}}| \mathbf{y}}\}$, then $R(D)=0$. The minimum is found with $d_m = \min\{\lambda,\lambda_m^{D}\}$, for $\lambda>0$ satisfying \eqref{eq:InProcConst} \cite{Cover:book}.

The achievability of the derived lower bound follows by considering the set of tuples in $\mathcal{R}_{\mathrm{IP}}(D)$ in Theorem \ref{th:inline} for $\mathbf{U}_l$ satisfying the additional Markov chain $\mathbf{U}_l \mkv \mathbf{R}_l \mkv (\mathbf{U}_{l-1}^*,\mathbf{S}_l) \mkv  \mathbf{S}_{l+1}$, which is included in $\mathcal{R}_{\mathrm{IP}}(D)$, as $\mathbf{U}^*_l=\mathbf{R}_l+\mathbf{Q}_l$, with $\mathbf{R}_l= \mathbf{P}^*_l[\mathbf{U}^*_{l-1},\mathbf{W}_l\mathbf{S}_{l}]^H$ and $\mathbf{Q}_l\sim \mathcal{CN}(\mathbf{0},\mathbf{K}^*_{l})$ and where $\mathbf{K}^*_l=\mathbf{V}_l\Lambda_l^{Q}\mathbf{V}_l^H$ and $\mathbf{P}^*_l=\mathbf{T}_{\mathbf{\bar{z}},\mathbf{f}}$. 
%
\qed

The distortion-rate function of the proposed IP scheme in Gaussian C-MIMO is given next. 
\begin{theorem}\label{lem:InlineGauss}
Given $\mathbf{U}_{\mathcal{L}}^*$ with $\mathbf{K}_1^*,\ldots, \mathbf{K}_L^*$ and $\mathbf{P}^*_1,\ldots, \mathbf{P}_L^*$ successively obtained as in Proposition \ref{th:RemoteProblem_inno}, the distortion-rate of the proposed IP scheme function is given as
\begin{align}\label{eq:InnoGaus}
D_{\mathrm{IP}}(R_1,\ldots,R_L)=&\mathrm{Tr}\{\mathbf{\Sigma}_{\mathbf{z}}-\mathbf{\Sigma}_{\mathbf{z},\mathbf{u}^*_{L}}\mathbf{\Sigma}_{\mathbf{u}^*_{L}}^{-1}\mathbf{\Sigma}_{\mathbf{z},\mathbf{u}^*_{L}}^H \},
\end{align}
where
$\mathbf{\Sigma}_{\mathbf{z},\mathbf{u}^*_L} = \mathbf{W}\mathbf{\Sigma}_{\mathbf{s}}\bar{\mathbf{P}}_{L}^{S,H}$;
$\mathbf{\Sigma}_{\mathbf{f}_{l}} = \mathbf{\bar{P}}_l^S\mathbf{\Sigma}_{\mathbf{s}}\bar{\mathbf{P}}_{l}^{S,H} + \mathbf{\bar{P}}_L^{Q}\text{diag}[\mathbf{K^*_{\mathcal{L}}}]\mathbf{\bar{P}}_L^{Q}$;
$\mathbf{\Sigma}_{\mathbf{u}^*_{L}} = \mathbf{\Sigma}_{\mathbf{f}_{L}} + \mathbf{K}^*_L$, 
and 
$ \mathbf{\Sigma}_{\mathbf{f}_l,\mathbf{s}_{l+1}} = \mathbf{\bar{P}}_l^{S}\mathbf{\Sigma}_{\mathbf{s}}\boldsymbol{\delta}_{l+1}$.
\end{theorem}
\text
\noindent \textbf{Proof:} 
Achievability follows from  Theorem \ref{th:inline} with $\mathbf{U}^*_{\mathcal{L}}$ obtained  as in Proposition \ref{th:RemoteProblem_inno}.
\qed

\begin{figure}
\centering 
\includegraphics[width=0.65\textwidth]{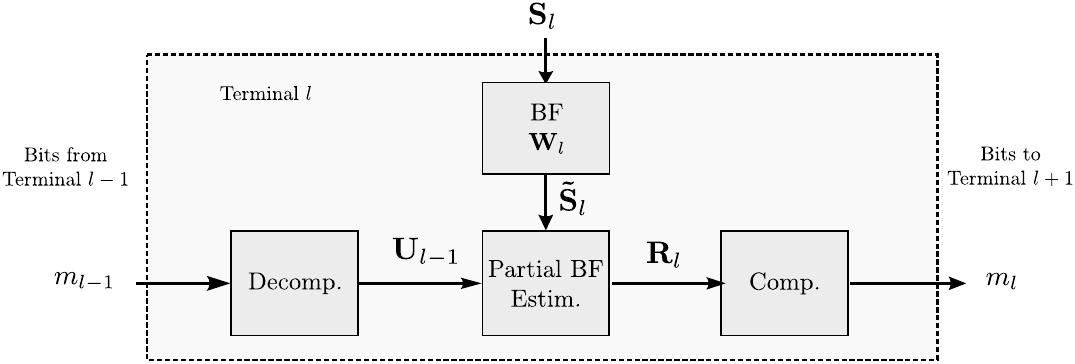}
\caption{In-network Processing scheme for C-MIMO.}\label{fig:Inline}
\vspace{-4mm}
\end{figure}

The IP scheme in Section \ref{ssec:InlineProcessing} requires joint compression at each RRU. However, for the Gaussian C-MIMO, it is shown next that the distortion-rate function $D_{\mathrm{IP}}(R_1,\ldots,R_L)$ in Theorem \ref{lem:InlineGauss} can be achieved by applying at each RRU separate decompression, partial function estimation followed by compression, as shown in Figure \ref{fig:Inline}.
At RRU $l$:
\begin{itemize}
\item Upon receiving $m_{l-1}$, decompress $\mathbf{U}_{l-1}$.
\item Apply local beamforming as $\tilde{\mathbf{S}}_l=\mathbf{W}_l\mathbf{S}_l$. 

\item Linearly combine $\mathbf{U}_{l-1}$, $\mathbf{\tilde{S}}_l$ to compute an estimate $\mathbf{R}_l= \mathbf{P}_l^*[\mathbf{U}_{l-1},\mathbf{\tilde{S}}_{l}]^H$ of the partial function up to Terminal $l$:
\begin{align}
\mathbf{Z}_l= \mathbf{W}_1 \mathbf{S}_1+\cdots+\mathbf{W}_l\mathbf{S}_{l},
\end{align}

\item Forward a compressed version of $\mathbf{R}_l$ to Terminal $(l+1)$ using Wyner-Ziv compression considering $\mathbf{S}_{l+1}$ as side information and the test channel $\mathbf{U}_l=\mathbf{R}_l+\mathbf{Q}_l$, $\mathbf{Q}_l\sim \mathcal{CN}(\mathbf{0},\mathbf{K}^*_{l})$. 
\end{itemize}

Terminal $(L+1)$ reconstructs $\mathbf{Z}$ using an MMSE estimator as $\mathbf{\hat{Z}}= \mathrm{E}[\mathbf{Z}|\mathbf{U}_L]$. 
\begin{proposition}\label{Prop:IPProp}
For the C-MIMO model, separate decompression,  partial function estimation and Wyner-Ziv compression achieves the distortion-rate function $D_{\mathrm{IP}}(R_1,\ldots, R_L)$ in Theorem \ref{lem:InlineGauss}.
\end{proposition}
\noindent \textbf{Proof:} 
The proof follows by showing that at any RRU $l$, the minimum distortion $D_l(R_l)$ and the test channel $\mathbf{U}^*_l$ in Proposition \ref{th:RemoteProblem_inno} can also be obtained with separate decompression, partial function estimation and compression. RRU $l$ decompresses $\mathbf{U}^*_{l-1}$ and computes $\tilde{\mathbf{S}}_l= \mathbf{W}_l\mathbf{S}_l$ and $\mathbf{R}_l = \mathbf{P}_l^*[\mathbf{U}_{l-1},\mathbf{\tilde{S}}_{l}]^H$. From standard arguments, it follows that compressing \`a-la  Wyner-Ziv with $\mathbf{S}_{l+1}$ as decoder side information requires
\begin{align}
R_{l}&\geq I(\mathbf{R}_l;\mathbf{U}_l|\mathbf{S}_{l+1})\\
& = I(\mathbf{V}_{l}^H\mathbf{R}_l;\mathbf{U}_l|\mathbf{S}_{l+1}) \label{eq:WZ_orthog} \\
& = I(\mathbf{V}^H\mathbf{T}_{\mathbf{\bar{z}_l},\mathbf{f}_l}\mathbf{F}_l;\mathbf{U}_l|\mathbf{S}_{l+1}),
\end{align}
where \eqref{eq:WZ_orthog} follows since $\mathbf{V}_l$ is orthonormal. 

Following from \eqref{eq:Achievable}, and by noting that the distortion achievable by estimating $\mathbf{\bar{Z}}_l$ from $\mathbf{\bar{U}}_{l}^*$ and $\mathbf{S}_{l+1}$ corresponds to $D_l$  in Proposition \ref{th:RemoteProblem_inno}, it follows that any achievable distortion $D_l$  is also achievable with separate decompression, partial function estimation and compression.
\qed

\section{A Lower Bound}\label{sec:Converse}

In this section, we obtain an outer bound on the RD region $\mathcal{R}(D)$ using a Wyner-Ziv type system in which the decoder is required to estimate the value of some function $Z$ of the input at the encoder $X$ and the side information $Y$ \cite{Yamamoto:IT:1982}. We use the following notation from \cite{Shamai:IT:98}.  Define the minimum average distortion for $Z$ given $Q$ as
$\mathcal{E}(Z|Q)\triangleq \min_{f:Q\rightarrow Z}\mathrm{E}[d(Z,f(Q))]$, and the Wyner-Ziv type RD function for value $Z$, encoder input  $X$ and side information $Y$ available at\mbox{ the decoder, as} \cite{Yamamoto:IT:1982}
\begin{align}\label{eq:WZFun}
R^{\mathrm{FWZ}}_{Z,X|Y}(D)\triangleq \min_{p(u|x):\mathcal{E}(Z|U,Y)\leq D} I(X;U|Y).
\end{align}
An outer bound can be obtained using the rate-distortion Wyner-Ziv type function in \eqref{eq:WZFun}.

\begin{theorem}\label{th:Converse}
The RD region $\mathcal{R}(D)$ is contained in the region $\mathcal{R}^{\mathrm{o}}(D)$, given by the union of tuples $(R_1,\ldots, R_{L})$ satisfying 
\begin{align}
R_l\geq R^{\mathrm{FWZ}}_{Z,S_{\mathcal{L}_l}|S_{\mathcal{L}_l^c}}(D),\quad  l=1,\ldots,L. 
\end{align}
\end{theorem}
\noindent \textbf{Outline Proof:} 
The outer bound is obtained by the RD region of $L$ network cuts, such that for the $l$-th cut, $S_{l+1}^n,\ldots,S_{L}^n$ acts as side information at the decoder. See Appendix \ref{app:Converse}.
\qed

In the Gaussian C-MIMO model, Theorem \ref{th:Converse} can be used to explicitly write a lower bound on the achievable distortion for a given fronthaul tuple $(R_1,\ldots, R_L)$ as given next.
\begin{proposition}
Given the fronthaul tuple $(R_1,\ldots,R_L)$, the achievable distortion in a C-MIMO system is lower bounded by 
$D_{\mathrm{LB}}(R_1,\ldots,R_L)= \max_{l=1,\ldots,L} D_l$,
where 
\begin{align}
D_l = \min_{d_{l,1},\ldots, d_{l,M}>0}\sum_{m=1}^{M}d_{l,m}\text{ s.t. } R_l = \sum_{m=1}^{M}\log^+\left(\frac{\lambda^{D}_{l,m}}{d_{l,m}}\right),
\end{align}
with $d_{l,m} = \min\{\lambda_l,\lambda_{l,m}^D\}$, for $\lambda_l>0$ and $\lambda^{D}_{l,m},m=1,\ldots, M$ are the eigenvalues of $\mathbf{\Sigma}_{\mathbf{z}|\mathbf{s}_{\mathcal{L}_l^c}}= \mathbf{W}\mathbf{\Sigma}_{\mathbf{s}|\dv s_{\mc L^c_l}} \mathbf{W}^H = \mathbf{V}_l\boldsymbol\Lambda_l^{D}\mathbf{V}^H_l$, with $\boldsymbol\Lambda_{l}^{D}=\text{diag}[\lambda^{D}_{l,1},\ldots, \lambda^{D}_{l,M}]$.
\end{proposition}
\noindent \textbf{Outline Proof:} 
The poof follows by computing explicitly the Wyner-Ziv RD type function in \eqref{eq:WZFun} for Gaussian vector sources for each network cut, which follows in the same lines as the proof of Proposition \ref{th:RemoteProblem_inno} with $\mathbf{F}=\mathbf{S}_{\mathcal{L}_l}$, $\mathbf{Y} = \mathbf{S}_{\mathcal{L}^c_l}$ and $\mathbf{Z} = \mathbf{W}\mathbf{S}$. Note that in this case $\mathbf{\Sigma}_{\mathbf{z}|\mathbf{f},\mathbf{y}} = \mathbf{0}$. 
\qed


\section{Numerical Results}\label{sec:NumRes}
In this section, we provide numerical examples to illustrate the average sum-distortion obtained using IR and IP as detailed in Section \ref{sec:Beamforming}.  We consider several C-MIMO examples, with $K$ users and $L$ RRUs, each equipped with $M$ antennas under different fronthaul capacities. The CP wants to reconstruct the receive-beamforming signal using the Zero-Forcing weights given by 
\begin{align}
\mathbf{W}  = (\mathbf{H}_{\mathcal{L}}^H\mathbf{H}_{\mathcal{L}})^{-1}\mathbf{H}_{\mathcal{L}}. 
\end{align}
The channel coefficients  are distributed as $h_{l,k}\sim\mathcal{CN}(0,1)$. We also consider the SR scheme of \cite{Park2015Multihop}. The schemes are compared among them, and to the lower bound in Theorem \ref{th:Converse}.  Note that WZR achieves the same distortion-rate function as IR as shown in Theorem \ref{lem:BTGaus}, \mbox{and is omitted.
}

\begin{figure}[!t]
\begin{minipage}[t]{0.5\linewidth}
\centering 
\includegraphics[scale=.6]{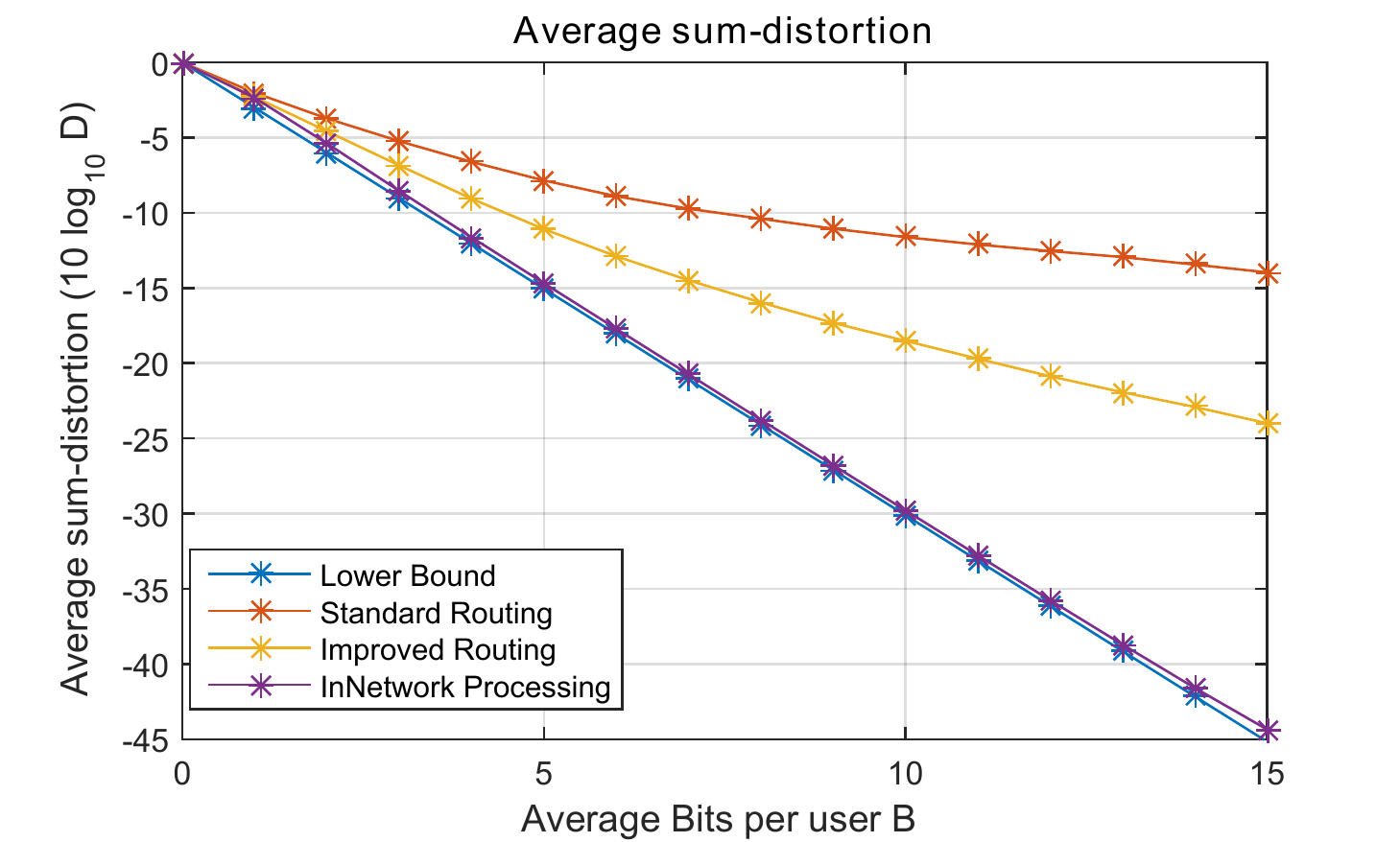}
\vspace{-6mm}
\caption{ Average sum-distortion $D$ for $M=15$, $L = 4$, $K =7$ vs. average bits per user for $B =0,\ldots, 15$ for balanced FH capacities $R_l = KB$.}\label{fig:Balanced_scaled}
\end{minipage}
\hspace{0.1cm}
\begin{minipage}[t]{0.5\linewidth} 
\centering 
\includegraphics[scale=.6]{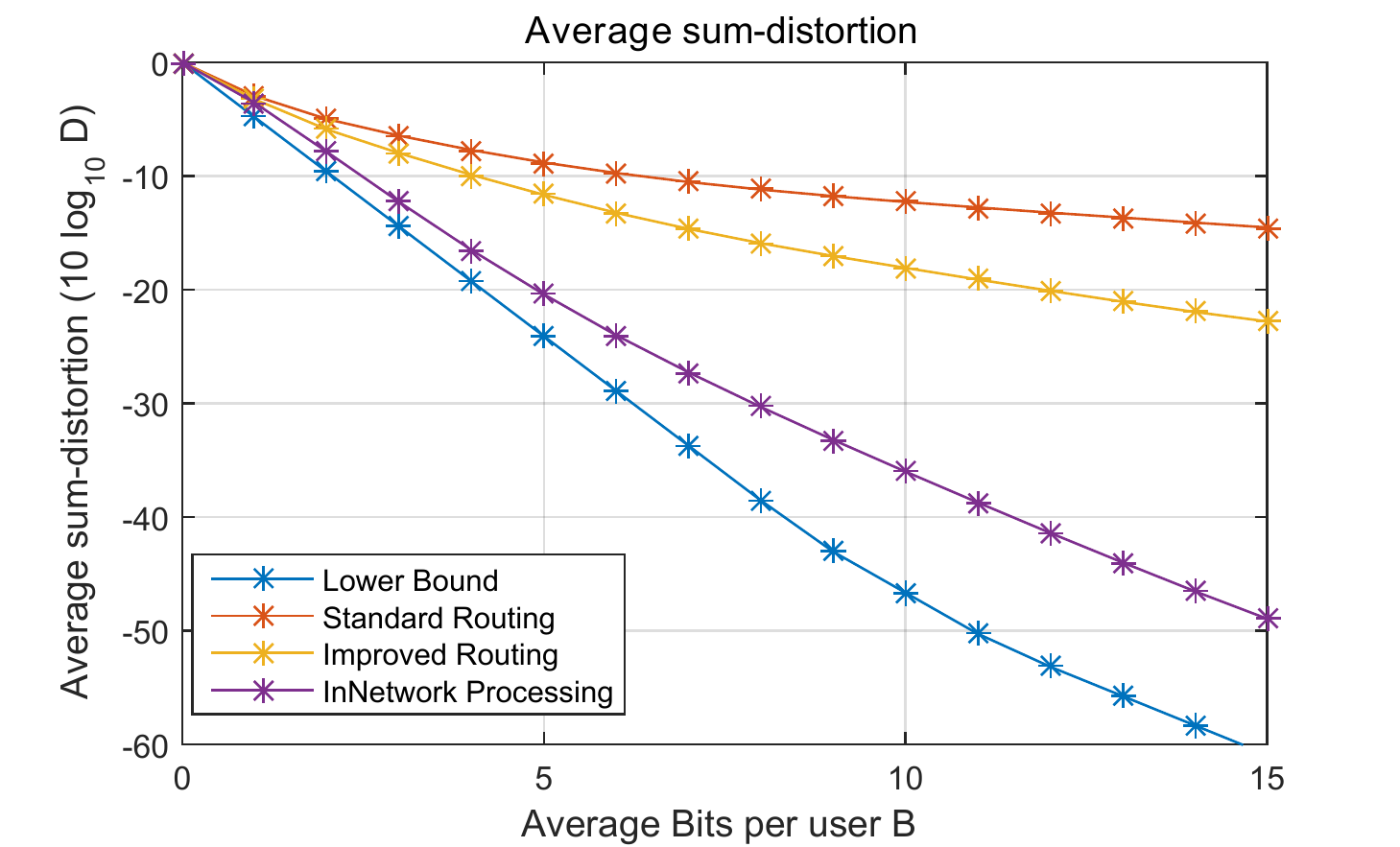}
\vspace{-6mm}
\caption{ Average sum-distortion $D$ for $M=15$, $L = 4$, $K =7$ vs. average bits per user for $B =0,\ldots, 15$ for increasing FH  $R_l = lKB$.}\label{fig:Unbalanced_scaled} 
\end{minipage}
\vspace{-6mm}        
\end{figure}

 
\noindent Figure \ref{fig:Balanced_scaled} depicts the sum-distortion in a C-MIMO network 
with $K = 15$ users and $L=4$ RRUs, each equipped with $M=7$ antennas for equal fronthaul capacity per link $R_1=\ldots = R_L = KB$,  as a function of the average number of bits per user $B$. As it can be seen from the figure, the scheme IP based on distributed beamforming outperforms the other centralized beamforming schemes, and performs close to the lower bound. For centralized beamforming, the scheme IF performs significantly better than SR, as it reduces the required fronthaul by only compressing the innovation at each RRU. 

Figure \ref{fig:Unbalanced_scaled}  shows the sum-distortion in a C-MIMO network with $K = 15$ users and $L=4$ RRUs, each equipped with $M=7$ antennas, with increasing fronthaul capacity per link $R_l = l KB$, $l=1,\ldots, L$ as a function of the average number of bits per user $B$. In this case, the IP scheme using distributed beamforming also achieves the lowest sum-distortion among the proposed schemes.

%
%
%

\bibliographystyle{ieeetran}
\bibliography{ref}

\begin{thebibliography}{10}
\providecommand{\url}[1]{#1}
\csname url@samestyle\endcsname
\providecommand{\newblock}{\relax}
\providecommand{\bibinfo}[2]{#2}
\providecommand{\BIBentrySTDinterwordspacing}{\spaceskip=0pt\relax}
\providecommand{\BIBentryALTinterwordstretchfactor}{4}
\providecommand{\BIBentryALTinterwordspacing}{\spaceskip=\fontdimen2\font plus
\BIBentryALTinterwordstretchfactor\fontdimen3\font minus
  \fontdimen4\font\relax}
\providecommand{\BIBforeignlanguage}[2]{{%
\expandafter\ifx\csname l@#1\endcsname\relax
\typeout{** WARNING: IEEEtran.bst: No hyphenation pattern has been}%
\typeout{** loaded for the language `#1'. Using the pattern for}%
\typeout{** the default language instead.}%
\else
\language=\csname l@#1\endcsname
\fi
#2}}
\providecommand{\BIBdecl}{\relax}
\BIBdecl

\bibitem{Estella:ICC:17:Chained}
I.~Estella and A.~Zaidi, ``In-network compression for multiterminal cascade
  mimo systems,'' in \emph{to appear in Proc. IEEE Int'l Conference on
  Communications (ICC)}, Paris, France, May 2017.

\bibitem{Cuff2009_Cascade}
P.~Cuff, H.~I. Su, and A.~E. Gamal, ``Cascade multiterminal source coding,'' in
  \emph{Proc. IEEE Int'l Symposium on Information Theory Proceedings (ISIT)},
  Jun. 2009, pp. 1199--1203.

\bibitem{Permu2012_CascTriang}
H.~Permuter and T.~Weissman, ``Cascade and triangular source coding with side
  information at the first two nodes,'' \emph{IEEE Tran. Inf. Theory}, vol.~58,
  no.~6, pp. 3339--3349, Jun. 2012.

\bibitem{Sefidgaran2016DistFunComp}
M.~Sefidgaran and A.~Tchamkerten, ``Distributed function computation over a
  rooted directed tree,'' \emph{IEEE Tran. Inf. Theory}, vol.~PP, no.~99, pp.
  1--1, Feb. 2016.

\bibitem{Park2015Multihop}
S.~H. Park, O.~Simeone, O.~Sahin, and S.~Shamai, ``Multihop backhaul
  compression for the uplink of cloud radio access networks,'' \emph{IEEE
  Trans. on Vehic. Tech.}, vol.~PP, no.~99, pp. 1--1, May 2015.

\bibitem{Gover2016ISIT}
Y.~Yang, P.~Grover, and S.~Kar, ``Coding for lossy function computation:
  Analyzing sequential function computation with distortion accumulation,'' in
  \emph{IEEE Int'l Symposium on Information Theory (ISIT)}, Jul. 2016, pp.
  140--144.

\bibitem{Puglielli:2015ICCW:ScalableMassive}
A.~Puglielli, N.~Narevsky, P.~Lu, T.~Courtade, G.~Wright, B.~Nikolic, and
  E.~Alon, ``A scalable massive {MIMO} array architecture based on common
  modules,'' in \emph{2015 IEEE Int'l Conference on Communication Workshop
  (ICCW)}, Jun. 2015, pp. 1310--1315.

\bibitem{Somekh:2007:IT}
O.~Somekh, B.~Zaidel, and S.~Shamai, ``Sum rate characterization of joint
  multiple cell-site processing,'' \emph{IEEE Trans. Inf. Theory}, vol.~53,
  no.~12, pp. 4473--4497, Dec. 2007.

\bibitem{DelCoso:2009:TWir}
A.~Del~Coso and S.~Simoens, ``Distributed compression for {MIMO} coordinated
  networks with a backhaul constraint,'' \emph{IEEE Trans. Wireless Comm.},
  vol.~8, no.~9, pp. 4698--4709, Sep. 2009.

\bibitem{Sanderovich:2009:IT}
A.~Sanderovich, O.~Somekh, H.~Poor, and S.~Shamai, ``Uplink macro diversity of
  limited backhaul cellular network,'' \emph{IEEE Trans. Inf. Theory}, vol.~55,
  no.~8, pp. 3457--3478, Aug. 2009.

\bibitem{Hwan:2013:VT}
S.-H. Park, O.~Simeone, O.~Sahin, and S.~Shamai, ``Robust and efficient
  distributed compression for cloud radio access networks,'' \emph{IEEE Trans.
  Vehicular Technology}, vol.~62, no.~2, pp. 692--703, Feb. 2013.

\bibitem{Park:2013:SPLett}
------, ``Joint decompression and decoding for cloud radio access networks,''
  \emph{IEEE Signal Processing Letters}, vol.~20, no.~5, pp. 503--506, May
  2013.

\bibitem{Zhou:2014:JSAC}
Y.~Zhou and W.~Yu, ``Optimized backhaul compression for uplink cloud radio
  access network,'' \emph{IEEE Journal on Sel. Areas in Comm.}, vol.~32, no.~6,
  pp. 1295--1307, Jun. 2014.

\bibitem{Nazer:ISIT:2009}
B.~Nazer, A.~Sanderovich, M.~Gastpar, and S.~Shamai, ``Structured superposition
  for backhaul constrained cellular uplink,'' in \emph{Proc. IEEE Int'l
  Symposium on Information Theory (ISIT)}, Seoul, Korea, Jun. 2012.

\bibitem{HongCaire:IT:2013}
S.-N. Hong and G.~Caire, ``Compute-and-forward strategies for cooperative
  distributed antenna systems,'' \emph{IEEE Trans. Inf. Theory}, vol.~59,
  no.~9, pp. 5227--5243, Sep. 2013.

\bibitem{Estella2016CQCF2}
I.~Estella-Aguerri and A.~Zaidi, ``Lossy compression for compute-and-forward in
  limited backhaul uplink multicell processing,'' \emph{IEEE Trans.
  Communications}, vol.~64, no.~12, pp. 5227--5238, Dec. 2016.

\bibitem{Estella:Allerton20015}
I.~Estella and A.~Zaidi, ``Partial compute-compress-and-forward for limited
  backhaul uplink multicell processing,'' in \emph{Proc. 53rd Annual Allerton
  Conf. on Comm., Control, and Computing}, Monticello, IL, Sep. 2015.

\bibitem{Shepard:2012:APM:2348543.2348553}
C.~Shepard, H.~Yu, N.~Anand, E.~Li, T.~Marzetta, R.~Yang, and L.~Zhong,
  ``Argos: Practical many-antenna base stations,'' in \emph{Proc. of the 18th
  Annual Int'l Conference on Mobile Computing and Networking (Mobicom '12)},
  2012, pp. 53--64.

\bibitem{Shepard:2013:AFM:2500423.2505302}
C.~Shepard, H.~Yu, and L.~Zhong, ``{ArgosV2:} a flexible many-antenna research
  platform,'' in \emph{Proc. of the 19th Annual International Conference on
  Mobile Computing and Networking (MobiCom '13)}.\hskip 1em plus 0.5em minus
  0.4em\relax New York, NY, USA: ACM, Sep. 2013, pp. 163--166.

\bibitem{Viera2014FlexibleTest}
J.~Vieira, S.~Malkowsky, K.~Nieman, Z.~Miers, N.~Kundargi, L.~Liu, I.~Wong,
  V.~\"{O}wall, O.~Edfors, and F.~Tufvesson, ``A flexible 100-antenna testbed
  for massive {MIMO},'' in \emph{2014 IEEE Globecom Workshops (GC Wkshps)},
  Dec. 2014, pp. 287--293.

\bibitem{Balan:2013:USE:2491246.2491254}
H.~V. Balan, M.~Segura, S.~Deora, A.~Michaloliakos, R.~Rogalin, K.~Psounis, and
  G.~Caire, ``{USC SDR}, an easy-to-program, high data rate, real time software
  radio platform,'' in \emph{Proc. of the Second Workshop on Software Radio
  Implementation Forum (SRIF '13)}, Aug. 2013, pp. 25--30.

\bibitem{Yamamoto:IT:1982}
H.~Yamamoto, ``{Wyner - Ziv} theory for a general function of the correlated
  sources (corresp.),'' \emph{IEEE Tran. Inf. Theory}, vol.~28, no.~5, pp.
  803--807, Sep 1982.

\bibitem{elGamal:book}
A.~E. Gamal and Y.-H. Kim, \emph{Network Information Theory}.\hskip 1em plus
  0.5em minus 0.4em\relax Cambridge University Press, 2011.

\bibitem{WZ76}
A.~D. Wyner and J.~Ziv, ``The rate-distortion function for source coding with
  side information at the decoder,'' \emph{{IEEE} Trans. Inf. Theory}, vol.~22,
  pp. 1--10, Jan. 1976.

\bibitem{Berger1989MultiterminalSourceEncoding}
T.~Berger and R.~Yeung, ``Multiterminal source encoding with one distortion
  criterion,'' \emph{IEEE Trans. Inf. Theory}, vol.~35, no.~2, pp. 228--236,
  Mar. 1989.

\bibitem{Cover:book}
T.~M. Cover and J.~A. Thomas, \emph{Elements of Information Theory}.\hskip 1em
  plus 0.5em minus 0.4em\relax Wiley-Interscience, 1991.

\bibitem{tian2009remote}
C.~Tian and J.~Chen, ``Remote vector {Gaussian} source coding with decoder side
  information under mutual information and distortion constraints,'' \emph{IEEE
  Tran. Inf. Theory}, vol.~55, no.~10, pp. 4676--4680, Oct. 2009.

\bibitem{Shamai:IT:98}
S.~Shamai, S.~Verd\'{u}, and R.~Zamir, ``Systematic lossy source-channel
  coding,'' \emph{IEEE Trans. Inf. Theory}, vol.~44, no.~2, pp. 564--579, Mar.
  1998.

\end{thebibliography}
\appendices

\section{Proof of Theorem \ref{th:Converse}}\label{app:Converse}
Suppose there exist $f_l^{(n)}$, $l=1,\ldots,L$ and $g^{(n)}$ such that for  $(R_1,\ldots, R_L)$, $\frac{1}{n}\sum_{i=1}^n\mathrm{E}[d(Z_i;\hat{Z}_i)]\leq D+\epsilon$, where $\epsilon \rightarrow 0$ as $n\rightarrow \infty$. Define $U_{l,i}\triangleq (m_l, S_{\mathcal{L}^c_l}^{-i}), $ for $l=1,\ldots, L$, where $S_{\mathcal{L}^c_l}^{-i}\triangleq (S_{\mathcal{L}_l^c,1}^{i-1},S_{\mathcal{L}_l^c,i+1}^{n})$ and note the Markov chain relation 
\begin{align}\label{eq:MK_chain}
S_{\mathcal{L}_{l}^c,i} \mkv S_{\mathcal{L}_{l},i} \mkv U_{l,i}, \quad l=1,\ldots, L.
\end{align}
\vspace{-3mm}
For the $l$-th cut we have, 
\begin{align}
nR_l&\geq H(m_l)\\
&\geq I(m_l;S^n_{\mathcal{L}}|S_{\mathcal{L}_{l}^c}^{n})\\
&=\sum_{i=1}^nH(S_{\mathcal{L},i}|S_{\mathcal{L}^c_{l},i})-H(S_{\mathcal{L}_l,i}|
S_{\mathcal{L}^c_{l}}^n, m_l,S_{\mathcal{L},1}^{i-1})\label{eq:conv_1}\\
&\geq\sum_{i=1}^nH(S_{\mathcal{L},i}|S_{\mathcal{L}^c_{l},i})-H(S_{\mathcal{L},i}|
S_{\mathcal{L}^c_{l},i}, m_l, S_{\mathcal{L}^c_l}^{-i})\label{eq:conv_2}\\
&=\sum_{i=1}^nI(S_{\mathcal{L},i};U_{l,i}|S_{\mathcal{L}^c_{l},i})\label{eq:data_sum}\\
&\geq \sum_{i=1}^nR^{\mathrm{FWZ}}_{Z,S_{\mathcal{L}_{l}}|S_{\mathcal{L}^c_{l}}}(\mathcal{E}(Z_{i}|U_{l,i},S_{\mathcal{L}^c_{l},i})),
\end{align}
where \eqref{eq:conv_1} follows since $S^n_{\mathcal{L}}$ is i.i.d., \eqref{eq:conv_2} follows since conditioning reduces entropy. On the other hand, we have
\begin{align}
\mathcal{E}(Z_{i}|U_{l,i},S_{\mathcal{L}^c_{l},i})
&= \mathcal{E}(Z_{i} |m_l, S_{\mathcal{L}^c_{l}}^n )\\
&= \mathcal{E}(Z_{i} |m_l,m_{l+1},\ldots,m_{L}, S_{\mathcal{L}^c_{l}}^n )\label{eq:dist_det_fun}\\
&\leq \mathcal{E}(Z_{i} |m_L)\label{eq:dist_det_fun_2}
\end{align}
where \eqref{eq:dist_det_fun} follows since $m_l$ is a deterministic function of $m_{l-1},S_l^n$, i.e., $m_l=f_l(m_{l-1},S_l^n)$, \eqref{eq:dist_det_fun_2} follows since reducing the information can only increase the distortion, 
Then, 
\begin{align}
nR_l
&\geq\sum_{i=1}^n R^{\mathrm{FWZ}}_{Z,S_{\mathcal{L}_{l}}|S_{\mathcal{L}^c_{l}}}(\mathcal{E}(Z_{i} |m_L))\label{eq:RWZ_monotonicity}\\
&\geq \sum_{i=1}^nR^{\mathrm{FWZ}}_{Z,S_{\mathcal{L}_{l}}|S_{\mathcal{L}^c_{l}}}(\mathrm{E}[d(Z_{i},\hat{Z}_i)]\label{eq:RWZ_monotonicity_2}\\
&\geq nR^{\mathrm{FWZ}}_{Z,S_{\mathcal{L}_{l}}|S_{\mathcal{L}^c_{l}}}(D+\epsilon),\label{eq:RWZ_monotonicity_3}
\end{align}
where \eqref{eq:RWZ_monotonicity} follows since $R^{\mathrm{FWZ}}_{Z,S_{\mathcal{L}_{l}}|S_{\mathcal{L}^c_{l}}}(D)$ is monotonic in $D$, \eqref{eq:RWZ_monotonicity_2} is due to $\hat{Z}_i$ being a function of $m_L$, and \eqref{eq:RWZ_monotonicity_3} follows as $R^{\mathrm{FWZ}}_{Z,S_{\mathcal{L}_{l}}|S_{\mathcal{L}^c_{l}}}(D)$ is convex and monotone in $D$.
This completes the proof.\qed

\end{document}